\documentclass[article]{revtex4}
\usepackage{amssymb,amsmath,amsfonts}
\usepackage{graphicx}

\newcommand\myfigwidth{0.66\linewidth}

\begin{document}

\title{Sheath parameters for non-Debye plasmas: simulations and arc damage}

\author{I. V. Morozov, G. E. Norman}
\affiliation{Joint Institute of High Temperatures of RAS, Moscow, Russia}

\author{ Z. Insepov, J. Norem$^*$}
\affiliation{Argonne National Laboratory, Argonne, IL 60439, USA}
 \email{norem@anl.gov}

\date{\today}

\begin{abstract}
This paper describes the surface environment of the dense plasma arcs that damage rf accelerators, tokamaks and other high gradient structures.   We simulate the dense, non-ideal plasma sheath near a metallic surface using Molecular Dynamics (MD) to evaluate sheaths in the non-Debye region for high density, low temperature plasmas.  We use direct two-component MD simulations where the interactions between all electrons and ions are computed explicitly.   We find that the non-Debye sheath can be extrapolated from the Debye sheath parameters with small corrections.  We find that these parameters are roughly consistent with previous PIC code estimates, pointing to densities in the range $10^{24} - 10^{25}\mathrm{m}^{-3}$.   The high surface fields implied by these results could produce field emission that would short the sheath and cause an instability in the time evolution of the arc, and this mechanism could limit the maximum density and surface field in the arc. These results also provide a way of understanding how the properties of the arc depend on the properties (sublimation energy, for example) of the metal.   Using these results, and equating surface tension and plasma pressure, it is possible to infer a range of plasma densities and sheath potentials from SEM images of arc damage.  We find that the high density plasma these results imply and the level of plasma pressure they would produce is consistent with arc damage on a scale 100 nm or less, in examples where the liquid metal would cool before this structure would be lost.  We find that the sub-micron component of arc damage, the burn voltage, and fluctuations in the visible light production of arcs may be the most direct indicators of the parameters of the dense plasma arc, and the most useful diagnostics of the mechanisms limiting gradients in accelerators.

\end{abstract}
\maketitle

\section{Introduction}

Comparatively little is known about the vacuum arcs and gradient limits that are important in determining the cost and overall parameters of large linear accelerator facilities.  Vacuum arcs are involved in many fields, from particle accelerators, plasma devices, high power switching, surface coating and a variety of laboratory and commercial applications, and these arcs have been under study for many years.  Nevertheless, the properties of these dense plasmas are not well understood, although the general behavior of these arcs has been known and under study since 1901 \cite{earhart, Anders,Juttner,mesyats,kajita} and these plasmas seem to limit the performance of both major accelerator and tokamak projects and facilities \cite{CLIC, MAP,ITER}.   In part, the reason for this situation is that the arcs are small, and many parameters (which are individually hard to measure) evolve very  rapidly over a very wide range.  Theory and modeling are complicated by the large number of mechanisms that seem to be involved in arc evolution and high density plasmas, which requires a non-Debye analysis of basic properties.

While arc damage has been measured and catalogued for over a hundred years, there has not been any clear correspondence between specific types of arc damage and the past or subsequent behavior of the arc.  We argue in this paper that the causes of arc damage are due to the high density, high surface electric field plasma that produces a high plasma pressure in the liquid metal surface underneath the plasma arc.  This pressure produces very small scale structures, at or below a few hundred nm.  We have developed a self consistent model of arc evolution and show that these structures may or may not survive the subsequent cool-down of the liquid surface, however the cool down itself also seems to produce cracks with small scale structures \cite{NoremPAC09, NoremLINAC10, Noremrf2011, InsepovNorem11}.  This damage can produce future breakdown events.

Recent work has shown that the development of the arc can be explained by two mechanisms:  1) mechanical failure of the solid surface due to Coulomb explosions caused by high surface fields \cite{InsepovNorem11}, and, 2) the development of unipolar arcs \cite{Schwirzke91}, that can act as virtual cathodes and produce currents that can short the driving potential.  Once an arc starts, the surface electric field, and field emission increase, increasing ionization of neutrals, causing an increase in the plasma density.  The density increase decreases the Debye length and causes an increase in the surface electric field, thus both the electric field and the density  increase exponentially with time, roughly described by the arrow in Fig. 1.  PIC simulations of the unipolar arc model for vacuum arcs relevant to rf cavity breakdown~\cite{NoremPAC09, NoremLINAC10, Noremrf2011, InsepovNorem11} show that the density of plasma formed above the field emitting asperities can be as high as $10^{26}\ \mathrm{m}^{-3}$. The temperature of such plasma is low, in the range of $1-10$~eV.

These high densities can make the Debye screening length,
\begin{align}
\lambda_D = \sqrt{\epsilon_0 k_B T / n_e e^2,}
\end{align}
become smaller than the mean inter-particle distance, or the number of particles in the Debye sphere,
\begin{align}
N_D = 4\pi n_e \lambda_D^3 / 3,
\end{align}
to become less that unity. This implies the failure of the ideal plasma approximation, as well as most of the assumptions used in simple calculations. Processes in such a dense plasma can be affected by three body particle collisions so that the Particle In Cell (PIC) method which relies on a simple collisional model, with two body collisions, becomes inappropriate, as shown in Fig 1, where the arrow shows the approximate range of parameters for evolving arcs from Ref  \cite{Noremrf2011,InsepovNorem11}, as well as the approximate region of validity for PIC  and Molecular Dynamics (MD) codes.

In this paper we calculate the parameters of the surface environment underneath the plasma sheath for the high density plasma conditions using direct two-component MD simulations where the interactions between all electrons and ions are computed explicitly. Although MD simulations have limited space and time scales their results can be considered as the lower level output for the multiscale approach.

Equilibrium and non-equilibrium nonideal plasmas have been studied
extensively by MD in the past several decades ~\cite{Zwicknagel-JPA03,JETP05,Murillo,Donko,Rostock-CPP09}.
Nevertheless there are few studies of the spatially inhomogeneous systems
such as electric double layers or plasma sheath. In this paper we report on
the first results for MD simulations of the nonideal plasma sheath near a
metallic surface.

We can compare modeling with experimental measurements, but both the modeling and experiments are somewhat indirect.  An understanding of the surface fields, combined with plasma density give estimates of the plasma pressure that can be compared with experimental data.  It is possible to make indirect experimental measurements of the plasma pressure by comparing the linear dimensions of structures seen in the surface with estimates obtained from comparing the surface tension of the liquid metal with the plasma properties.  We describe how the scale of structures frozen in damage as the arc cools, can be used to set limits on the plasma properties.

\section{Simulation technique}

The two-component fully ionized electron-ion plasma is considered. Neutral
atoms are not taken into account which can affect relaxation times at
relatively low plasma densities when the density of neutrals is high
enough. It should not, however, affect the stationary distribution of
charges. In the present work the simulations are restricted to the singly
ionized plasma with $Z=1$.

The electron-electron and ion-ion interactions are given by the Coulomb
potential. For electrons and ions it is modified at short distances to
account for quantum effects. The equation below assumes a Gaussian wave
function for an electron
\begin{align}
  V_{ei} (r) =  \frac{Z e^2}{4 \pi \epsilon_0 r}
    \mathrm{erf} \left( \frac{r}{r_0} \right), \label{erf}
\end{align}
where the $r_0$ parameter that equals to $0.21$~nm in out case to match the
ionization energy for copper at $r=0$: $U(0) = - 7.73$~eV (see
Fig.~\ref{potential-ei}). The similar interaction model was used e.g.
in~\cite{Rostock-CPP09,Suraud06,Zwicknagel-CPP03} for simulations of
ionized metallic clusters. More accurate electron-ion and electron-electron
interaction models are discussed e.g. in~\cite{CKelbg,WPMD-JPA09} although
they seem to be redundant for this particular case. In fact the results are
weakly dependent on the short distance part of the potential as the change
of the $U(0)$ value from 7.73~eV to 5.1~eV does not change the results
within simulation accuracy.

The Leap-Frog integration scheme is used to solve the classical equations
of motion for electrons and ions \cite{velocityVerlet}. The method takes into account the conservation of the total energy of the finite system, as long as there is
no external potential. To follow the electron dynamics, time steps of
$0.001-0.01$~fs were taken to calculate the time evolution.

The general simulation scheme follows the method described in
\cite{MolSimul05} and shown in Fig.~\ref{averaging}. First an equilibrium
MD trajectory is calculated for the system at given density and temperature
using the nearest image method (periodic boundary conditions) for all
dimensions. The simulation box size and other parameters are summarized in
Table~\ref{SimParam}. The Langevin thermostat~\cite{LangevinTherm} is used initially to bring the
system to an equilibrium state while it is switched off for a production
run. Then the system becomes adiabatic which ensures that all thermodynamic quantities are conserved in average. The ion mass is set to be equal to the electron mass for better mixing of ionic trajectories at this phase.  The nonideality parameter, $\Gamma$, is the ratio of the average Coulomb potential energy and the average kinetic energy per electron~\protect\cite{EKKR_book}.

\begin{table}[h!]
\begin{center}
\begin{tabular}{|c|c|c|c|c|c|c|c|}
\hline
$T, \mathrm{eV}$ & $n_e, 10^{27}\mathrm{m}^{-3}$ &
$L_x, \mathrm{nm}$ & $L_z, \mathrm{nm}$& $N_i$ &
$\Gamma$ & $\Theta$ & $\lambda_D, \mathrm{nm}$ \\
\hline
1  & 0.0001 & 120 & 360 & 518 & 0.11 & 0.001 & 23.5 \\
1  & 0.001  & 55  & 165 & 499 & 0.23 & 0.004 & 7.43 \\
1  & 0.01   & 25  & 75  & 468 & 0.50 & 0.017 & 2.35 \\
1  & 0.1    & 11  & 33  & 399 & 1.08 & 0.079 & 0.74 \\
1  & 1.0    & 5   & 15  & 375 & 2.32 & 0.36  & 0.24 \\
1  & 5.0    & 2.8 & 8.4 & 329 & 3.97 & 1.07  & 0.11 \\
10 & 0.01   & 25  & 75  & 468 & 0.05 & 0.002 & 7.43 \\
10 & 1.0    & 5   & 15  & 375 & 0.23 & 0.036 & 0.74 \\
10 & 100    & 1   & 2.5 & 300 & 1.08 & 0.79  & 0.07 \\
\hline
\end{tabular}
\caption{\label{SimParam}MD simulation parameters: $T$ is the initial
electron temperature, $n_e$ is the initial number density of electrons (or
ions), $L_z$ is the transversal simulation box size, $L_x$ is the
longitudinal simulation box size, $N_i$ is the number of ions which is
equal to the initial number of electrons, $\Gamma = e^2(4\pi
n_e/3)^{1/3}/(4\pi\epsilon_0 k_B T)$ is the nonideality parameter, $\Theta
= \hbar^2 (3\pi^2 n_e)^{2/3} / (2m_e k_B T)$ is the degeneracy parameter,
$\lambda_D$ is the Debye length.}
\end{center}
\end{table}

At the second phase the particle positions and velocities at particular
time moments are taken from the equilibrium trajectory to be used as the
initial states for nonequilibrium simulations of the plasma sheath. The
interval between those points should be large enough for the initial states
to be statistically independent from the microscopical point of view.
However, all these states correspond to the same macroscopical conditions
as they are taken from a single equilibrium trajectory. Then a bunch of
trajectories is computed starting from the given ensemble of initial states
and the results are averaged over the ensemble.

In order to study the plasma sheath, the XY plane at $z = 0$ axis is
considered as a metallic surface whereas a reflecting wall is introduced on
the other side of the box at $z = L_z$. The periodic boundary conditions
are still applied for transverse dimensions $x$ and $y$. When an electron
crosses the surface it is always meant to be absorbed. Therefore it is
removed from the system and the overall surface charge is incremented by
its charge $q_{\mathrm{surf}} \leftarrow q_{\mathrm{surf}} - e$.

A non-zero surface charge produces an electrostatic field which influence
the particles. where $\sigma = q_{\mathrm{surf}}/(L_x L_y)$ is the surface charge density and $L_x$, $L_y$ are the box sizes in the transverse dimensions.  Assuming $L_x = L_y = 2a$ one can
obtain (see e.g.~\cite{Jackson_book})
\begin{align}
   E_z(z)
    = \frac{\sigma}{\pi\epsilon_0} \arctan\left[ \frac{a^2}{z\sqrt{2a^2 + z^2}} \right],
  \label{Ez}
\end{align}
where $E_z$ is the longitudinal component of the electric field. It can be
shown that Eq.~(\ref{Ez}) tends to the field expression of a uniformly charged plane
$E_z = \sigma/(2\epsilon_0)$ as $z\to 0$ and to the Coulomb field $E_z =
\sigma a^2 / (\pi\epsilon_0) / z$ as $z\to\infty$. It is important to use
Eq.~(\ref{Ez}) in the simulation with the given boundaries instead of the
field of a charged plane $E_z = \sigma/(2\epsilon_0)$ as the later cannot be
screened by plasma particles at a large distance. As the surface field
grows it starts to repel electrons from the surface until the stationary state
is reached.

We do not compute dynamics of ions at this phase as the ions are too heavy
to contribute to the simulation results at the electron time scale. At the
same time the ions are movable at the equilibrium trajectory that is used
for generation of the initial states. Thus the averaging over an ensemble
means the averaging over different configurations of ions.   In a real system the number of ions will vary with time, due to ions entering and leaving the plasma from their thermal motion and self sputtering.  Because the ion velocity is 340 times smaller than the electron velocity, this process is very slow, and we have neglected these effects.  This is equivalent to assuming that the self sputtering coefficient for copper ions is near unity.  

We have checked that the final results are independent of the simulation box
size. If the box is doubled the deviation of the results are in within the
statistical errors.

The thermodynamics parameters was maintained in the course of simulation.
It was found that the overall electron temperature deviates in the range of
$1-10$\% due to absorption of the most energetic electrons to the surface.

\section{Simulation Results and Fit Formulas}

Typically the relaxation of the electric field is observed for about 1~ps
(see Fig.~\ref{surfield-t}). The development of the electron profile is
shown in Fig.~\ref{dp-t}. The stationary density profiles obtained after
the relaxation are shown in Fig.~\ref{dp-z}. As the ions does not move,
their distribution mimics the uniform distribution obtained from the
equilibrium trajectory with full periodic boundary conditions. On the
contrary, electrons form the well pronounced layer of plasma near the
surface with a positive charge which we consider as the plasma sheath.

The plasma charge density profile is given by the difference between the
ion and electron densities as presented in Fig.~\ref{charge-z} in the
semilogarithmic scale. It is seen that starting from the surface the charge
density $\sigma(z)$ decays exponentially which agrees with the Debye
approximation. At high densities, however, the exponential decay is
preceded by a non-exponential area. This
regions makes difference between calculation of the sheath length from the
slope $\lambda_\mathrm{exp}$ of the exponential decay
$\sigma(z)\sim\mathrm{e}^{-z/\lambda_\mathrm{exp}}$ and from the distance
$\lambda$ at which the charge density decreases at the value of $\mathrm{e}
= 2.71$ ($\sigma(\lambda) = \sigma(0)/\mathrm{e}$) as illustrated in
Fig.~\ref{charge-z}c.

Both quantities $\lambda_\mathrm{exp}$ and $\lambda$ are presented in
Fig.~\ref{lambda-n} depending on the plasma density and temperature. It is
seen that $\lambda_\mathrm{exp}$ follows the Debye-like dependence on
density $\lambda_\mathrm{exp} \sim n_e^{-1/2}$ whereas the real sheath
length $\lambda$ scales with a slightly different exponent.

The fits for MD data are
\begin{align}
  T_e = 1 \mathrm{eV}:\qquad &
  \lambda_\mathrm{exp} = 1.7\lambda_D, \quad
    \lambda[\mathrm{nm}] = 1.0\cdot10^{11} (n_e[m^{-3}])^{-0.405}, \label{lambdaT1}\\
  T_e = 10 \mathrm{eV}:\qquad &
  \lambda_\mathrm{exp} = 1.7\lambda_D, \quad
    \lambda[\mathrm{nm}] = 3.18\cdot10^{12} (n_e[m^{-3}])^{-0.449}. \label{lambdaT10}
\end{align}

Fig.~\ref{lambdar-n} shows how the ratios $\lambda_\mathrm{exp}/\lambda_D$ and $\lambda/\lambda_D$, depend on plasma temperature, electron density and the nonideality parameter.  While Fig.~\ref{lambdar-n}a shows the dependence of the screening length on plasma temperature,  Fig.~\ref{lambdar-n}b shows that the values of $\lambda/\lambda_D$ for different temperatures are close to each other when plotted versus the parameter $\Gamma$. This implies that the  nonexponential charge density decay and the difference between $\lambda$ and $\lambda_D$ are primarily a function of the plasma nonideality, defined as the ratio of average electrostatic potential energy divided by average kinetic energy.

The values of the electric field at the metal surface are presented in
Fig.~\ref{surfiled-n} depending on both temperature and density. The solid
line corresponds to Eq.~(2) from~\cite{Schwirzke91}
\begin{align} \label{EfSchwirzke}
  E = \frac{V_f}{\lambda_D} = [n_e k_B T_e / (4\epsilon_0)]^{1/2}
  \log[M_i / (2\pi m_e)],
\end{align}
where $M_i$ and $m_e$ are the masses of electron and ion. If the Debye
radius in Eq.~(\ref{EfSchwirzke}) is substituted by the MD
result~(\ref{lambdaT1}) or~(\ref{lambdaT10}) it results in the values shown
by crosses in Fig.~\ref{surfiled-n} which are in a better agreement with
the MD results.

The fits for MD data shown by dashed lines are
\begin{align}
  T_e = 1\mathrm{eV}:\qquad &
    E[\mathrm{GV/m}] = 2.57\cdot10^{-15} (n_e[m^{-3}])^{-0.577}, \label{EfT1}\\
  T_e = 10\mathrm{eV}:\qquad &
    E[\mathrm{GV/m}] = 1.21\cdot10^{-13} (n_e[m^{-3}])^{-0.531}. \label{EfT10}
\end{align}

Fig.~\ref{plasmapot-n} shows the plasma potential calculated using the
simple relation of $\phi = E / \lambda$ where both the surface electric
field $E$ and the sheath length $\lambda$ are obtained from MD simulations.
A more rigorous result can be found by integration of the electric field
distribution in plasma but it requires a more accurate evaluation of the
space charge away from the sheath area and will be the subject of future
work.

\section{Experimental estimates of plasma parameters}

The internal parameters of the arc, the surface field and the metal under it, are not directly accessible, but experimental measurements can provide some indirect evidence of the internal structure and the active mechanisms.   Three phenomena are, in principle, sensitive to these numbers: 1) the mechanism that limits the exponential increase of the density and electric field with time, 2) the properties of the metal surface (melting point, self sputtering yield, etc.) may determine the burn voltage through the mechanism of self sputtering, and, 3) the scale of surface damage frozen into the surface should be sensitive to the plasma pressure that created it.  We discuss these briefly.

\subsection{Field Emission Driven Plasma Instabilities}

The surface fields and plasma densities described above can be very high and it has been shown that these fields and densities increase exponentially with time during the evolution of the discharge \cite{NoremLINAC10, Noremrf2011, InsepovNorem11}.  The magnitude of these fields suggests that field emission over the entire active area could short out the sheath, causing an instability or oscillation in the plasma limiting this exponential increase.

As the surface field increases above 2 GV/m, it becomes possible for field emission currents to short out (or significantly reduce) the sheath field in times on the scale of nanoseconds.  Assuming this occurs, we would expect that the sheath would rapidly reestablish itself due to the short collision time, the comparatively large plasma volume, and the high plasma density, and this behavior would produce fluctuations in the visible emission of the arc and fluctuations in the thickness of the sheath.  This phenomenon could be described as the plasma "bouncing" on the metal surface.  It is known that arcs are unstable, and fluctuations of this sort have been described by J\"{u}ttner \cite{Juttner} and Anders, see Fig 3.22 of Ref. \cite{Anders}.  In rf accelerator cavities, we see oscillations in visible light detected by a phototube with a frequency of approximately 200 MHz that could be due to this effect, see Fig ~\ref{PMT} and Fig ~\ref{FFT}.

We can understand the parameter range involved by estimating the current density required to short the sheath in a given time.
\begin{align}
i_s = \Delta \sigma / \Delta t_s = \epsilon_0 E,
\end{align}
where $\Delta t$ = 1 ns implies currents of about 20 MA/m$^2$ and $\sigma$ is the surface charge density. We can approximate Fowler-Nordheim field emission expression for current density at low fields  \cite{dolan} with,
\begin{align}
i_{FN} = 1.8 \times 10^7 (E/3 \times 10^9)^{16},
\end{align}
where the current density $i$ is expressed in A/m$^2$ and the electric field is in V/m.  Thus, when electric fields of 3 GV/m appear on the surface they will produce field emission currents capable of shorting the sheath in 1 ns.  Assuming the plasma takes a few ns to return to the original density this would imply an instability with a time constant of a few ns.

Instabilities in arc evolution are a well studied phenomenon.  These strong oscillations may be related to the "ecton" model Mesyats has described, where he assumes micro-explosions with a timescale of $10^{-8}$ s \cite{mesyats}.  The characteristic, discontinuous "chicken track" traces on the interior of tokamak cavities could also be driven by these instabilities \cite{Anders,kajita}.  These instabilities are the subject of a further study, and will be be reported elsewhere.

We note that the field emission current densities discussed here, multiplied by the areas of melted copper in 805 MHz cavity damage spots, on the order of 2.5 $\times 10^{-7}$ m$^2$, would produce currents on the order of 4 A, roughly equal to the shorting current expected in 805 MHz breakdown events, see Fig 1 of Ref \cite{InsepovNorem11}.

\subsection{Material Dependence}

We expect some dependence of the sheath potential and sheath parameters on the properties of the surface material. It has been shown that the evolution of the plasma is primarily driven by field emitted electron beams at high electric field and self-sputtering of surface material driven by ions falling through the sheath potential \cite{NoremLINAC10, Noremrf2011, InsepovNorem11}. Self-sputtering produces a flux of neutral atoms that can raise the plasma density and also  fuel the plasma, permitting long plasma lifetimes \cite{kajita}.  Numerical simulations of sputtering yields show that this mechanism is very sensitive to the sheath potential, ion charge distribution, surface (melting) temperature \cite{NIMBselfsput}, and even grain orientation \cite{NIMBselfsput1}.

Anders, in Section 3.7 of Ref \cite{Anders} explains the "burn voltage" of arcs, {\it i.e.} the voltage drop at the cathode,  in terms of the Cohesive Energy Rule, where the cohesive energy of the cathode material is essentially the binding energy of the surface atoms.  The larger the cohesive energy, the larger the burn voltage (proportional to the sheath potential) required to maintain a plasma. (The burn voltage is related to the sheath potential, but not equal to it, since electrons emitted from the plasma are not necessarily emitted at the plasma potential.)  We believe that the mechanism that correlates the cohesive energy to the burn voltage is likely to be self-sputtering, which is determined by interatomic bonding.  This data suggests that the sheath potential should be related to the atomic bonding energy, since we have shown that the sheath potential is primarily a function of the plasma temperature and weakly dependent on the density (see Fig.~\ref{plasmapot-n}), we assume that this the plasma temperature is primarily involved.  Measurements relating the burn voltage to plasma temperature (and perhaps crystal orientation) in different materials might explore this correlation.

\subsection{Plasma Pressure and Surface Damage}

We believe that the nature of the surface damage can provide information on the parameters of the sheath and the arc.  The plasma ion flux hitting the surface should rapidly melt the top few layers of the surface. The plasma pressure pushing on the liquid metal surface can generate a Tonks-Frankel like instability \cite{He}, and uneven surfaces produced by this instability will be opposed by the surface tension force, which will tend to flatten the surface.  As the dimensions of this instability become smaller, the surface tension force becomes more dominant, producing a correlation between the plasma pressure and the spatial scale of damage.  The experimental problem is that surface tension will tend to smooth over the whole melted area when the liquid surface cools, making the melted areas polished and smooth.  Our approach is to look for the smallest scale structure visible in arc damage, and assume that cooling has been rapid enough to preserve some evidence of the plasma pressure.  For the scale of damage we observe ( $\sim 0.2 \times 10^{-6}$ m) thermal decay times are on the order of,
\begin{align}
 x = \sqrt{Dt},
 \end{align}
where $D$ is the thermal diffusivity constant, approximately $1.1 \times10^{-4}$ m$^2$/sec, implying times on the order a few ns or less, depending very strongly on the geometry of both the material and the details of the heat pulsing. We find experimentally that arcs moving in a transverse magnetic field produce the most fine structure.

The pressure exerted on the surface by the plasma is due to: a) the plasma, and b) the electric field.  The thermal plasma pressure is due to ion impacts, $p_i = nkT$, and the electric field pressure is defined by, $p_E = \epsilon_0 E^2/2$ \cite{Jackson_book}, the total is then, $p = p_i  - p_E$, since the ions push on the surface and the electric field, generally much smaller, pulls.  In the limit of small $E$ and $T$, the ion pressure can be a function primarily of the sheath potential, $p_i=ne\phi$.  If, due to a variety of reasons, the pressure is unevenly applied, it will produce a deformation in the liquid surface that is opposed by surface tension, see Fig ~\ref{p-gamma}.  The approximate scale of these effects is set by the equilibrium radius, $r$,  where the radius where the surface tension is balanced by the plasma pressure can obtained by equating the surface tension force around the circumference with the pressure over the whole area of a hemispherical bubble \cite{Shortley},
\begin{align}
2 \pi r \gamma = \pi r^2 p,
\end{align}
with $\gamma$ equal to the surface tension constant, approximately 1 N/m, depending on temperature, giving $r \sim 2\gamma/p$ \cite{copper}.  For small structures it has been shown by Tolman that this expression should be corrected by a factor $\delta$, using the expression,
\begin{align}
r = \frac{2\gamma}{p}  \left( 1-\frac{\delta}{r} + \cdots \right),
\end{align}
where $\delta$ is the Tolman length \cite{Tolman49}.  The Tolman length is generally evaluated using Molecular Dynamics, and estimates vary from tenths of molecular dimensions to hundredths of atomic dimensions.  For radii, $r$, on the order of 100 nm this correction is not significant.

There are many types of arc damage that have been seen SEM images \cite{Anders,kajita,castano}.  The damage from a single event is generally circular, in the range of 5 - 200 $\mu$m in diameter, and frequently craterlike with a raised rim.  The damage usually shows signs of melting.  If the surface has absorbed significant energy, fine structure from the arc can be lost as the metal solidifies,  However, if the arc deposits little energy to the surface or cools quickly, for example in SEM images of damage, Fig ~\ref{SEM}, a) from 201 MHz rf coupler, and b) from arc damage from Castano \cite{castano} and images from laser damage \cite{Schwirzke91},  we find complex structures on the scale of 100 - 300 nm, which are not seen in arc damage where large amounts of energy ($\sim$1 J) were present.   We assume that if large amounts of energy are transmitted through an arc crater there is less small scale structure, consistent with high stored heat keeping the metal liquid until the surface tension smoothed off the surface.  Classic unipolar arc tracks \cite{kajita} (where magnetic fields move the arc in rapid retrograde motion) are associated with more fine structure, consistent with faster liquid cool down preserving this fine structure.

Simulations of unipolar arcs using PIC codes \cite{InsepovNorem11} have shown that the plasma potential seems to stay approximately 50 to 75 V during the development of the arc, thus the variation in plasma pressure is primarily due to the plasma ion density.  Schwirzke showed that unipolar arcs could produce holes 5 times deeper than their diameter (~0.7 $\mu$m) \cite{Schwirzke91}.  If we assume that these structures grew from craters with $r \leq$ 0.2 - 0.35 $\mu$m, and the plasma potential, $\phi$ was 50 V, this would imply that the density of the plasma had to be at least $1 - 4 \times 10^{24}$  m$^{-3}$, see Fig ~\ref{r-n}.  This is consistent with estimates made from the PIC code, which would not be expected to be reliable at these high densities.

The primary arc damage that results in further high enhancement factors and further breakdown events is likely due to this sub-micron damage, coming either from the plasma pressure itself producing a turbulent surface if it can quickly cool, or cracks produced when the large molten area beneath the arcs cools from the melting point of copper to room temperature leaving a network of surface cracks.  The production of high enhancement factors in surface cracks has been demonstrated in Ref \cite{NoremLINAC10,Noremrf2011}.  The sub-micron component of arc damage thus appears to be both the most direct indicator of the internal parameters of the arc plasma, and (when cracks and crack junctions are considered) the most likely to produce further breakdown events due to high enhancement factors.

\section{Conclusions}

We used molecular dynamics simulations to study the non-ideal plasma sheath at a metal surface for conditions we believe are appropriate to those found in accelerator cavities or unipolar arcs. The simulations started from the uniform equilibrium plasma state. Then the relaxation of the electron density profile with formation of the plasma sheath was observed. The relaxation time was found to be of the order of $\sim$ 100 fs. It was shown that the plasma sheath length depends on the electron number density in a slightly different way than the usual expression for the Debye radius due to a non-exponential charge profile at short distances. The values of the sheath length and the surface field were obtained for two values of temperatures and a wide density range with the non-ideality parameter $\Gamma = 0.1 - 4$. We compare the MD results  with the contemporary theoretical models and with experimental data from damage. When we compare the plasma conditions that would result from these sheaths with data we find damage consistent with the high plasma pressures implied by the MD and PIC results.

We find that the high density plasma these results imply and the level of plasma pressure they would produce is consistent with the spatial scale of arc damage in rf cavities, in examples where the arc would cool before this structure would be lost.  It appears that the sub-micron component of arc damage is both the most direct indicator of the internal parameters of the arc plasma, and, in the case of cracks, the most likely source of further breakdown events due to high enhancement factors.  The high surface fields implied by these results could produce field emission that would short the sheath and cause an instability in the time structure of the arc. The relation between self sputtering and the burn voltage is not well understood but the two seem to be closely correlated.  We find that the sub-micron component of arc damage, the burn voltage, and fluctuations in the visible light production of arcs may be the most direct indicators of the sheath parameters of the dense plasma.

\section{Acknowledgements}

We  thank the staff of the Accelerator and Technical Divisions at Fermilab and the Muon Accelerator Program (MAP) for supporting and maintaining the MAP experimental program in the MTA experimental area.    The work at Argonne is supported by the U.S. Department of Energy Office of High Energy Physics  under Contract No. DE-AC02-06CH11357.  I. Morozov acknowledges the support by the Programs of Fundamental Research of RAS Nos. 2, 13 and 14. Computations were performed on clusters MIPT-60 and K100 (KIAM RAS).

\newpage

\begin{figure}[h]  
\begin{center}
  \includegraphics[width=\myfigwidth]{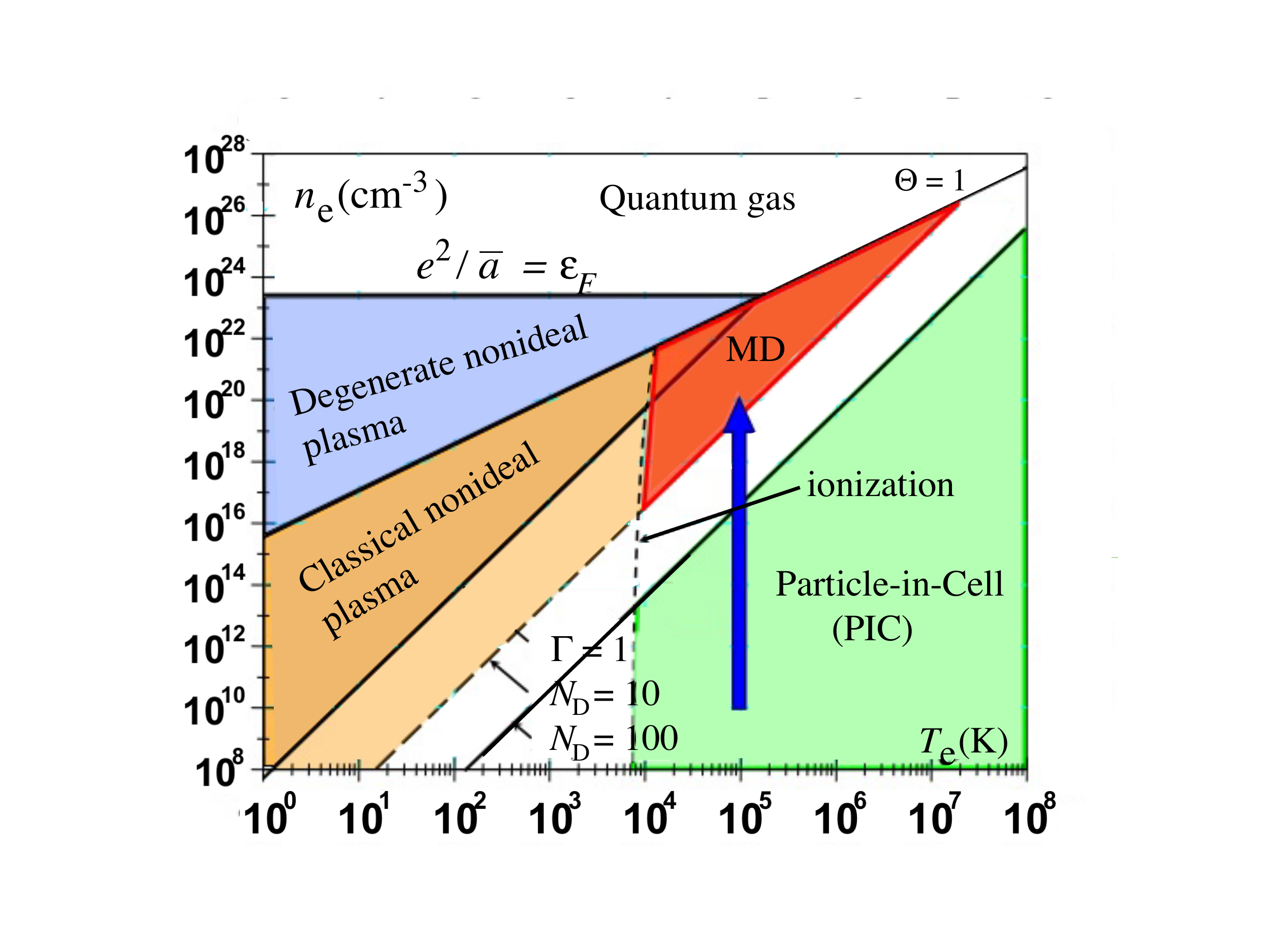}
  \vspace*{-14pt}
\end{center}
\caption{\label{PIC-MD} The range of PIC and MD codes.  The arrow shows the time development on an arc, as described in ref \cite{NoremPAC09, NoremLINAC10, Noremrf2011, InsepovNorem11} }
\end{figure}

\begin{figure}[h]  
\begin{center}
  \includegraphics[width=\myfigwidth]{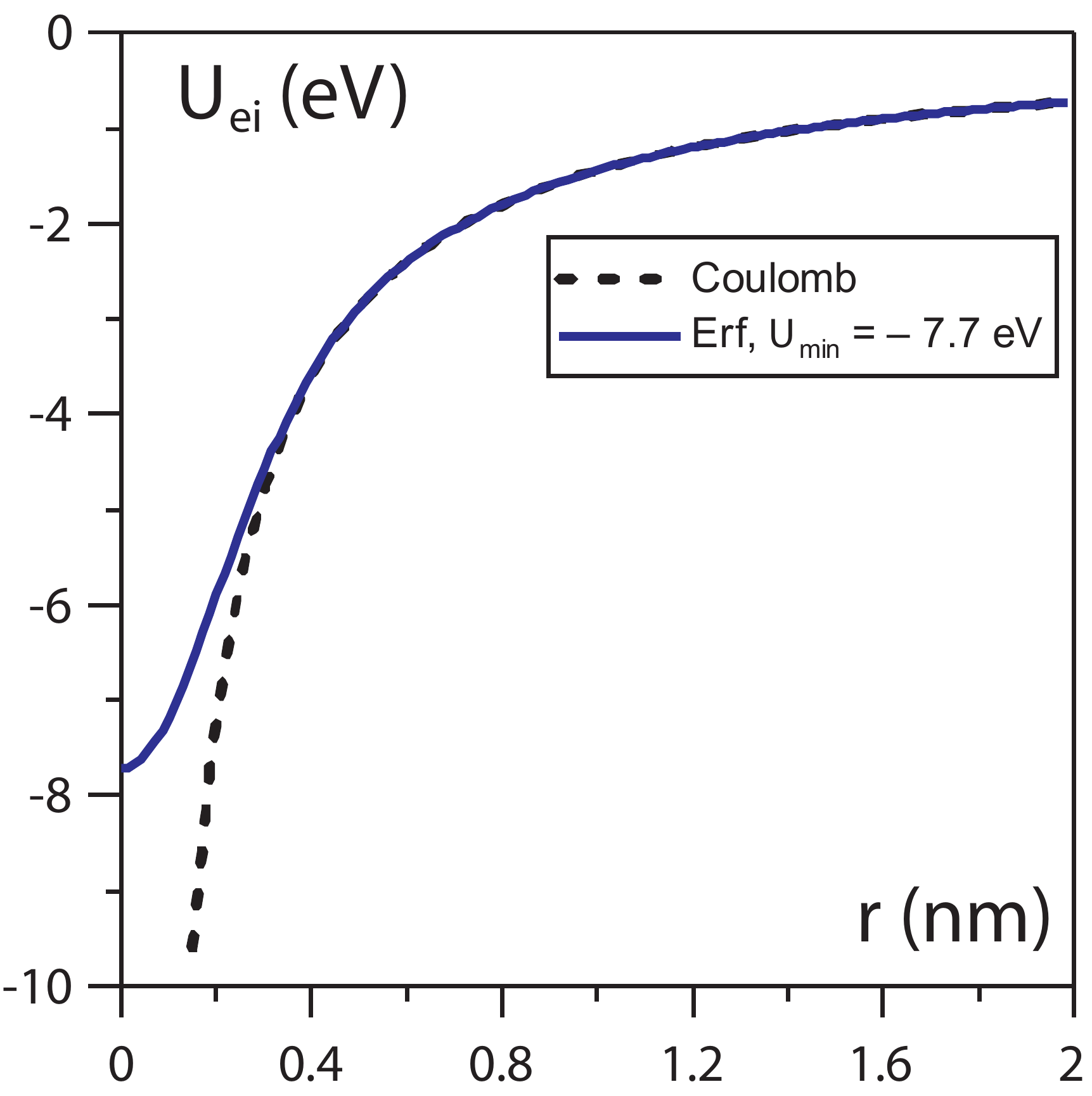}
  \vspace*{-14pt}
\end{center}
\caption{\label{potential-ei}Electron-ion interaction potential: dashed line -- pure Coulomb,
solid line -- the one used in this work.}
\end{figure}

\begin{figure}[h]  
\begin{center}
  \includegraphics[width=\myfigwidth]{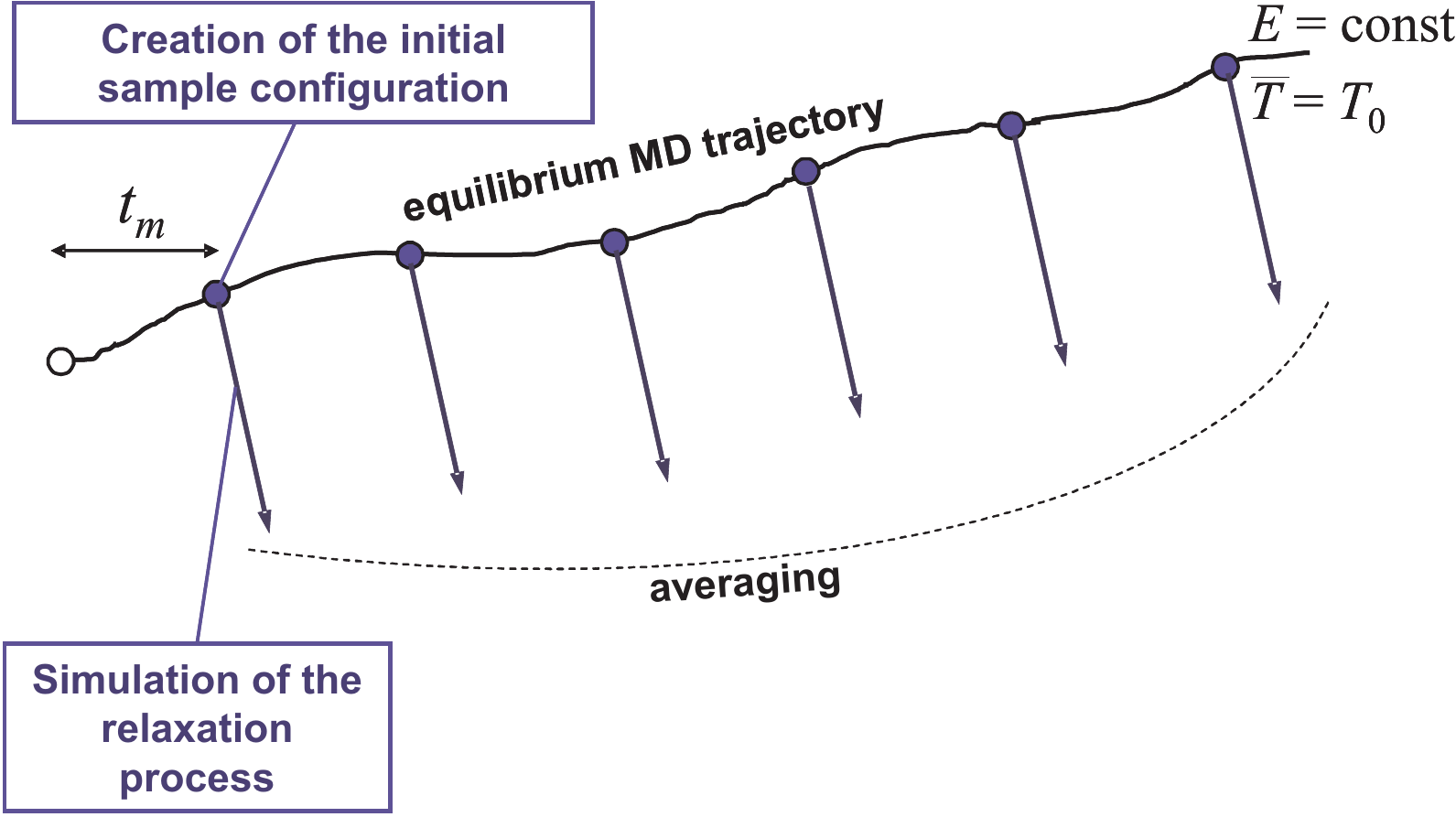}
  \vspace*{-14pt}
\end{center}
\caption{\label{averaging}General simulation scheme: averaging over an ensemble of initial states taken from
an auxiliary equilibrium trajectory (solid curve with points).}
\end{figure}

\begin{figure}[h]  
\begin{center}
  \includegraphics[width=\myfigwidth]{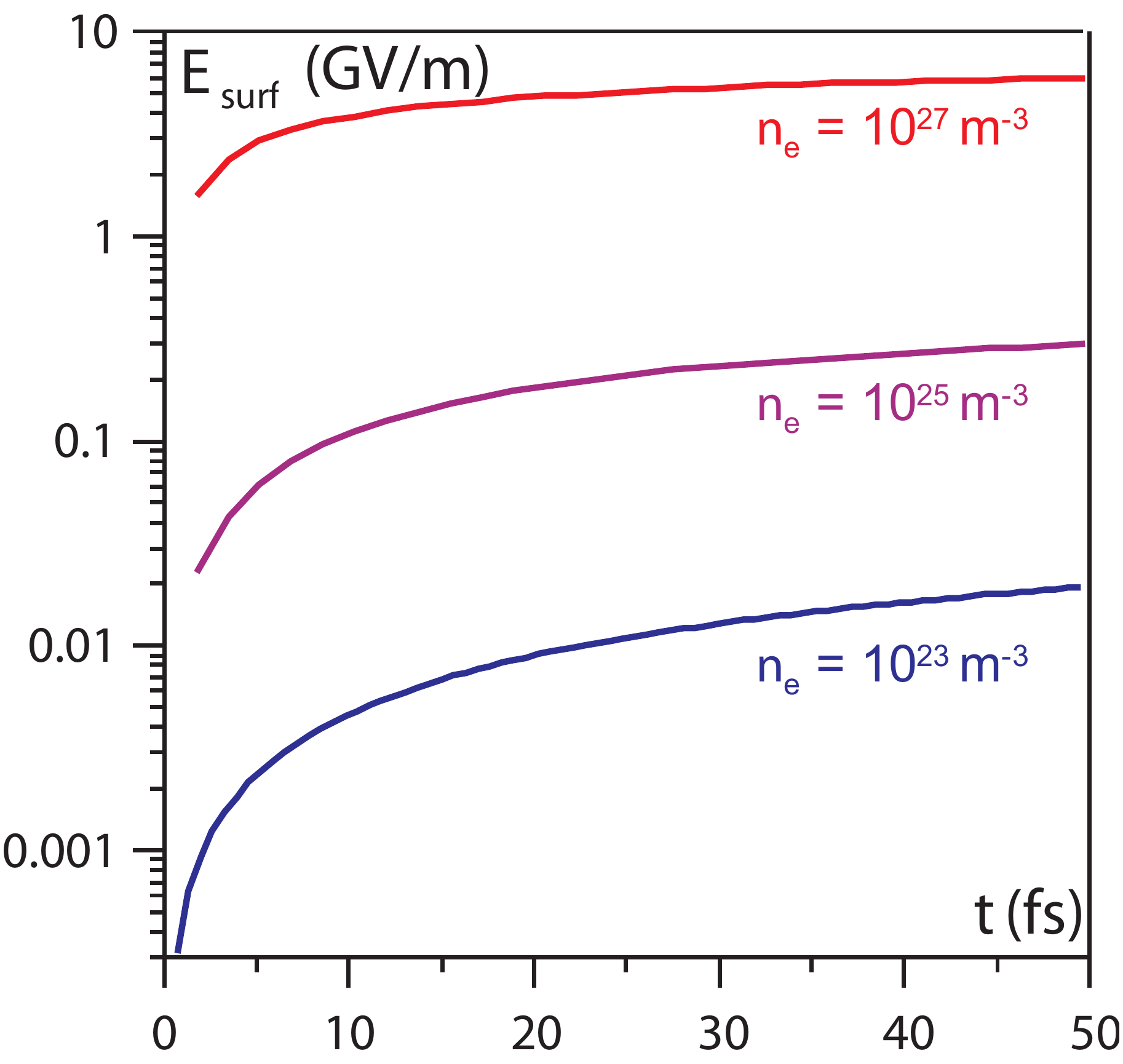}
  \vspace*{-14pt}
\end{center}
\caption{\label{surfield-t}Dependence of the electric field strength at the surface on time
for different electron number densities (shown on the plot); $T = 1$eV.}
\end{figure}

\begin{figure}[h]  
\begin{center}
  \includegraphics[width=\myfigwidth]{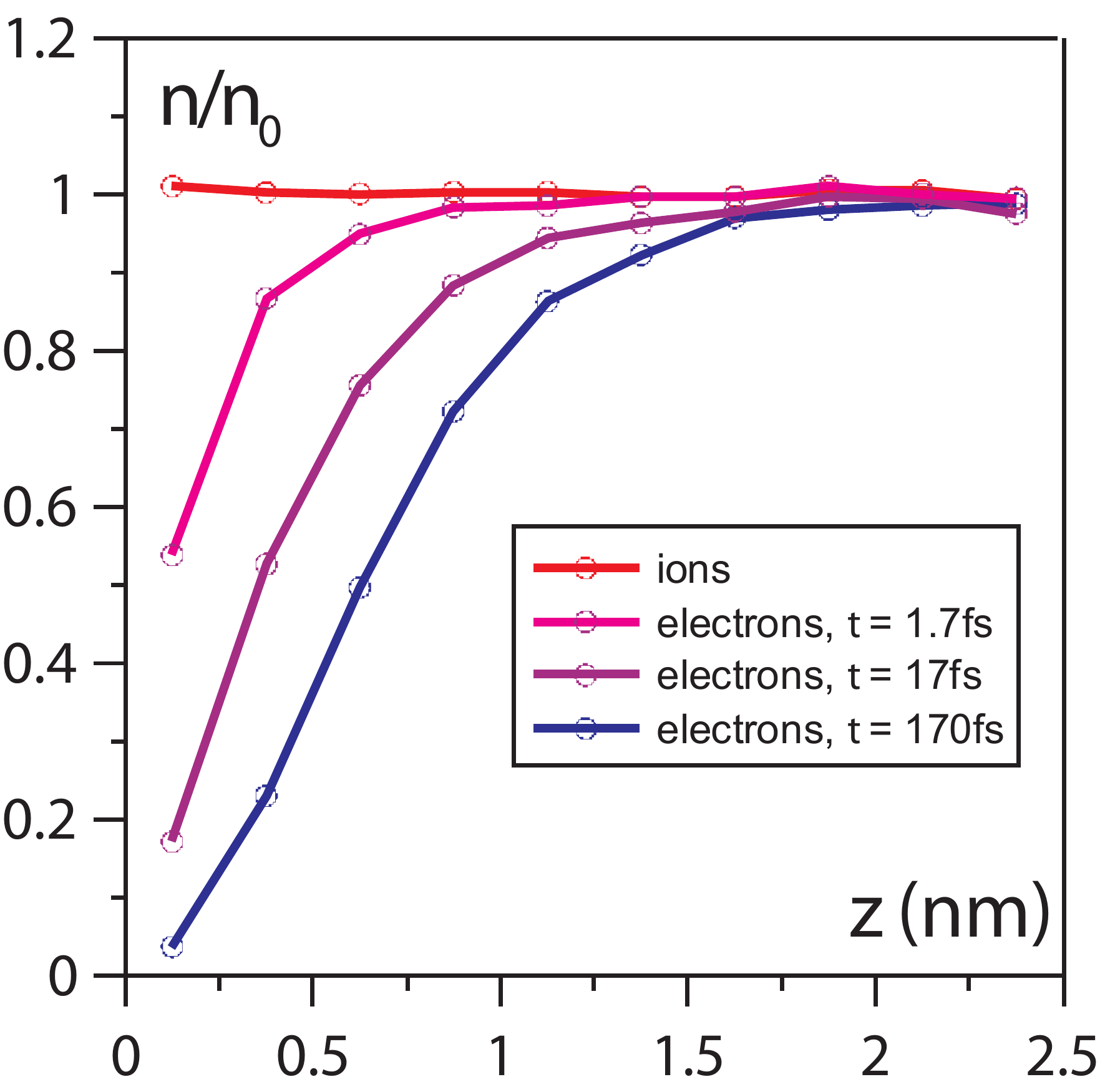}
  \vspace*{-14pt}
\end{center}
\caption{\label{dp-t}Development of the electron density profile with time.
The time moments are shown in the legend. The density is normalized by the mean density
in the initial state; $n_e = 10^{23}\mathrm{m}^{-3}$, $T = 1$eV.}
\end{figure}

\begin{figure}[h]  
\begin{center}
  \includegraphics[width=0.335\linewidth]{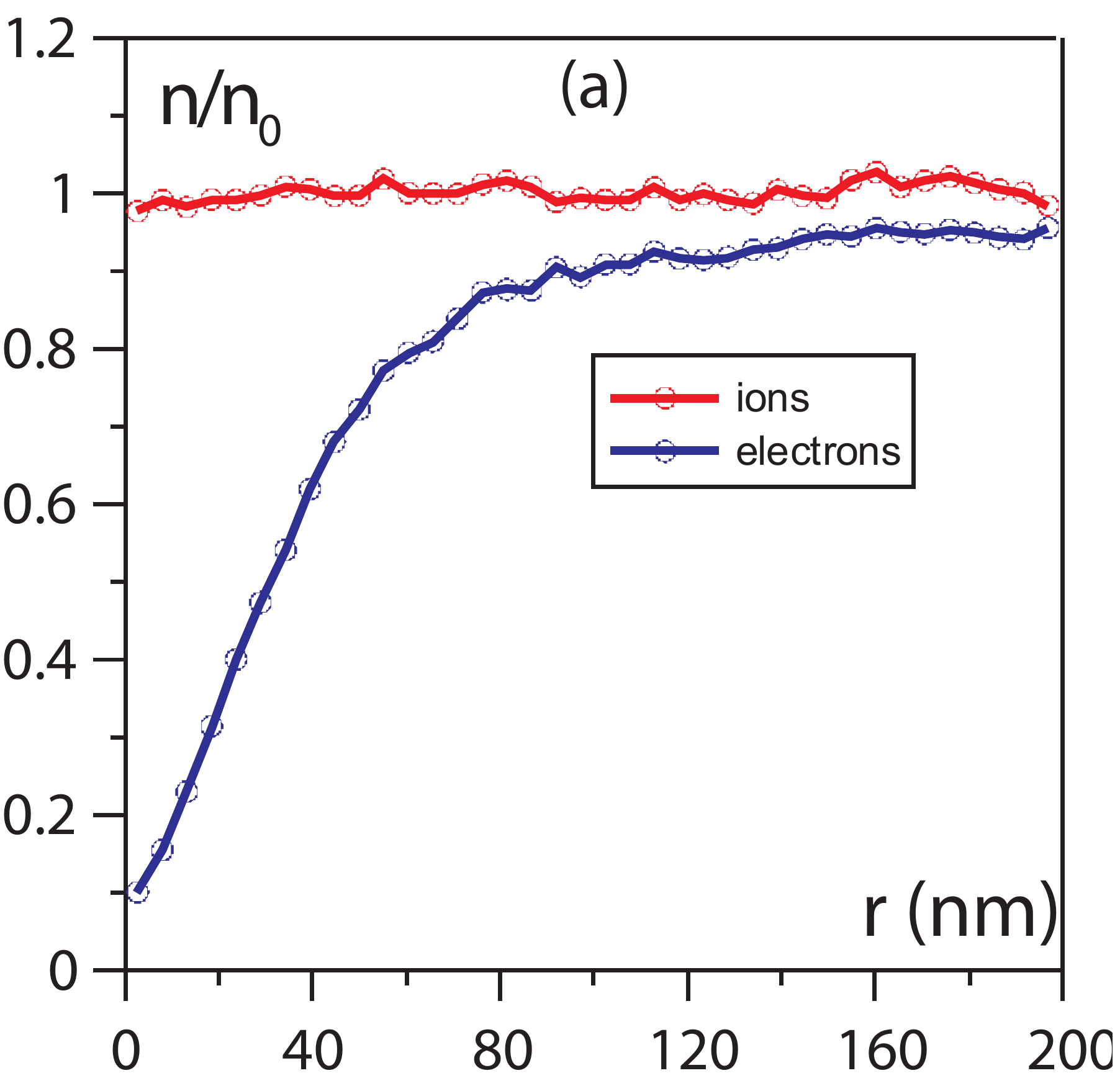}
  \includegraphics[width=0.303\linewidth]{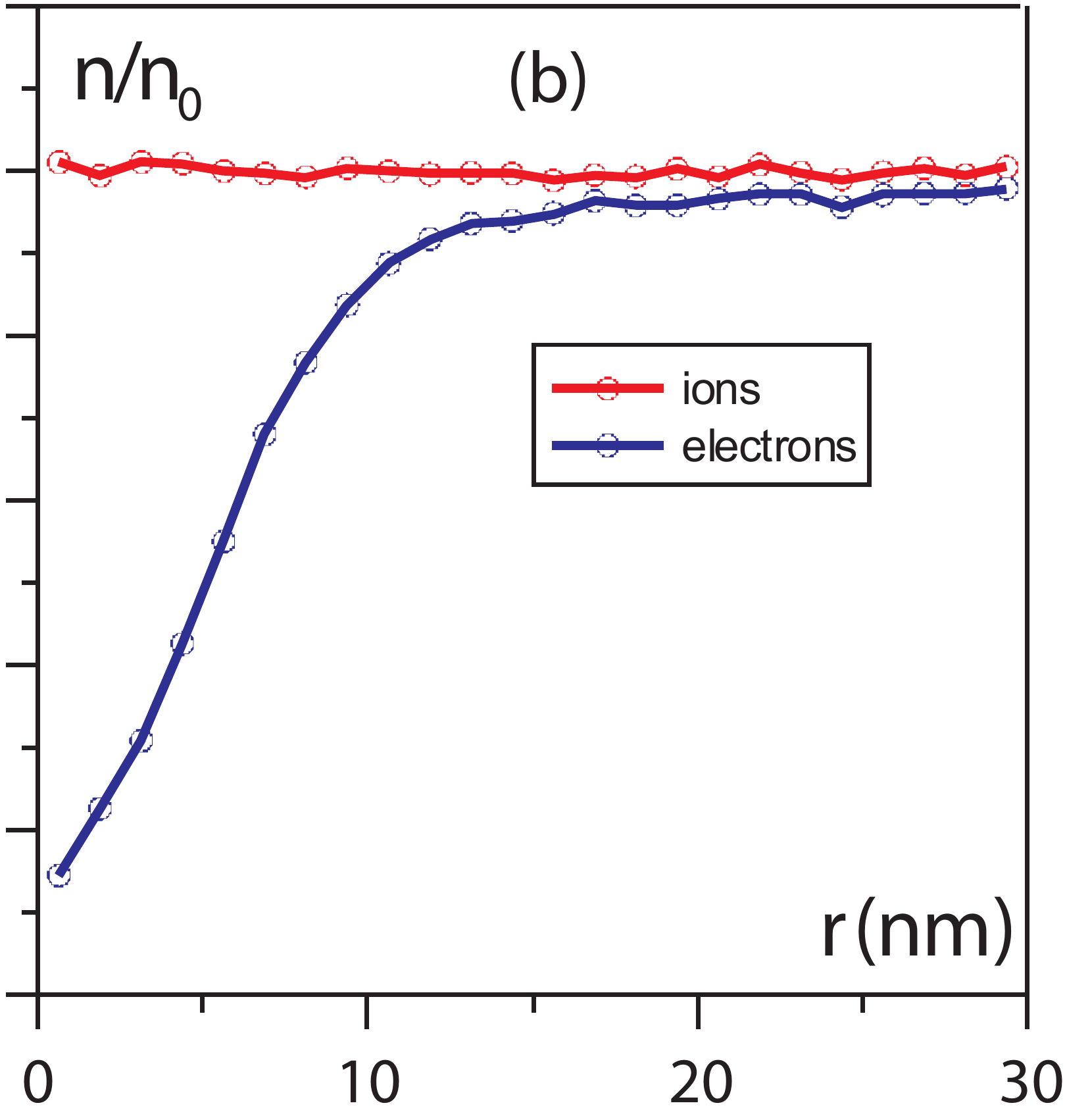}
  \includegraphics[width=0.30\linewidth]{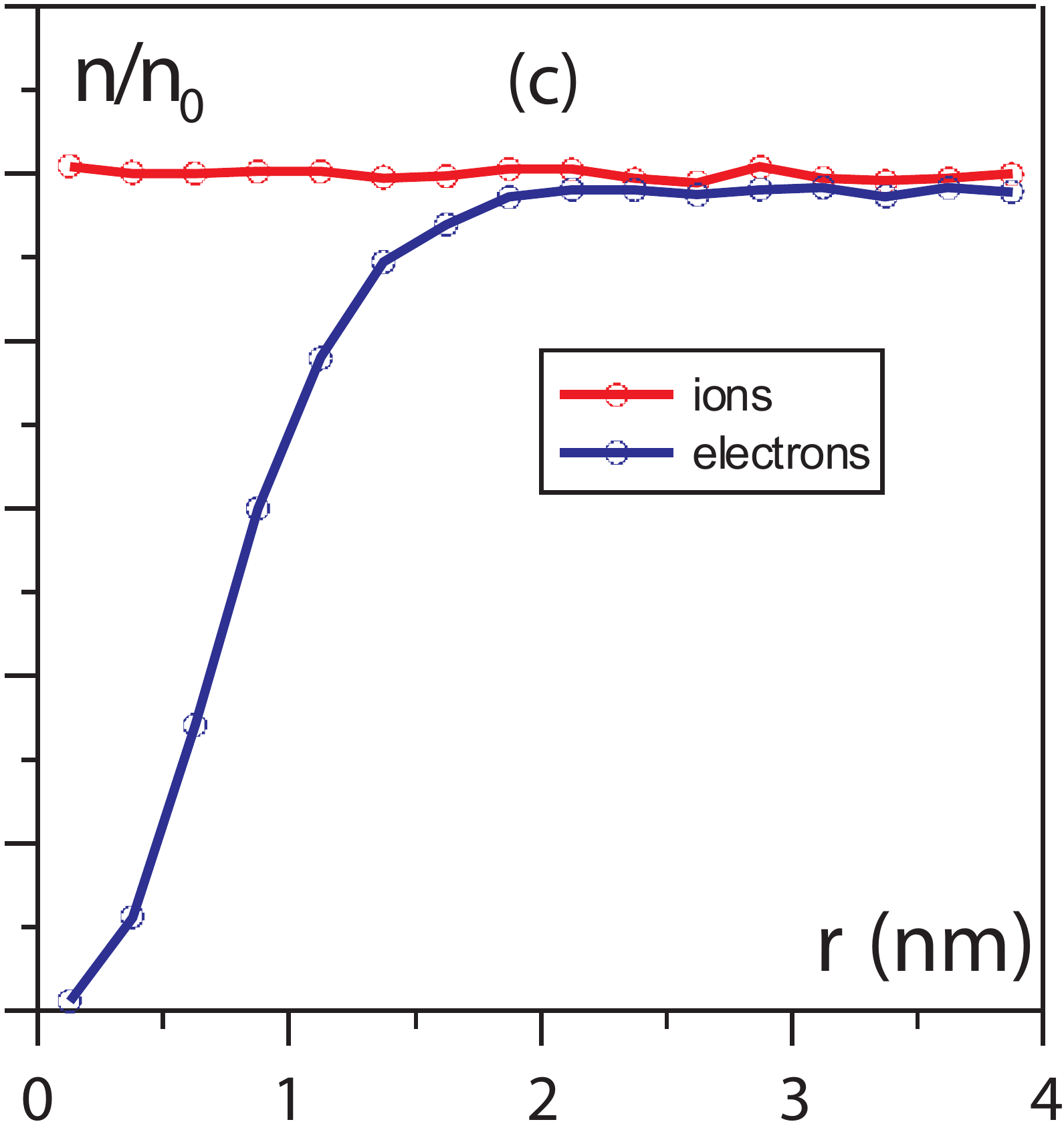}
  \vspace*{-14pt}
\end{center}
\caption{\label{dp-z}Density profiles for electrons and ions at the final stationary state
depending on the mean electron number density: (a) $n_e = 10^{23}\mathrm{m}^{-3}$
(b) $n_e = 10^{25}\mathrm{m}^{-3}$, (c) $n_e = 10^{27}\mathrm{m}^{-3}$. In all cases $T = 1$eV.}
\end{figure}

\begin{figure}[h]    
\begin{center}
  \includegraphics[width=0.325\linewidth]{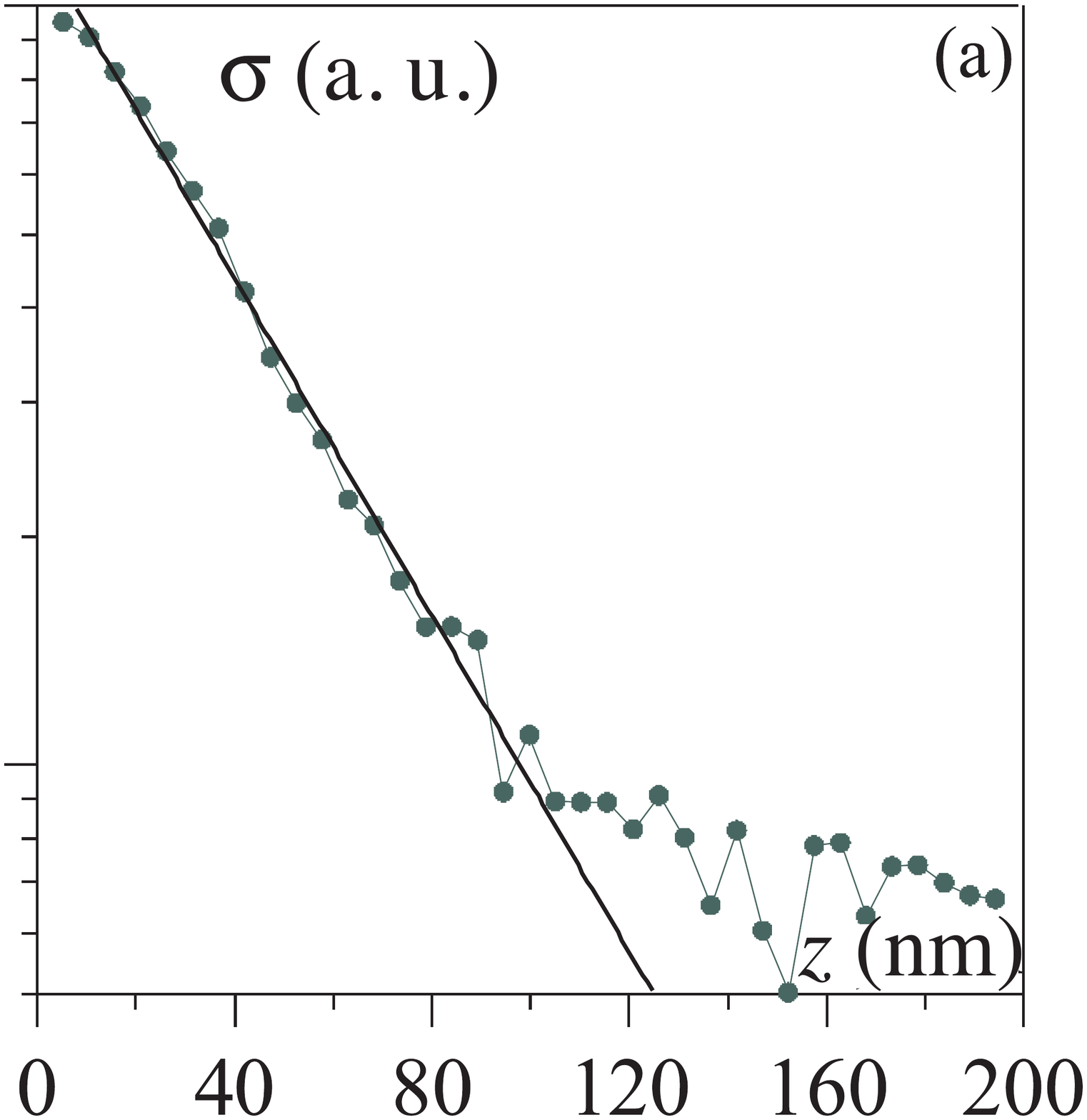}
  \includegraphics[width=0.322\linewidth]{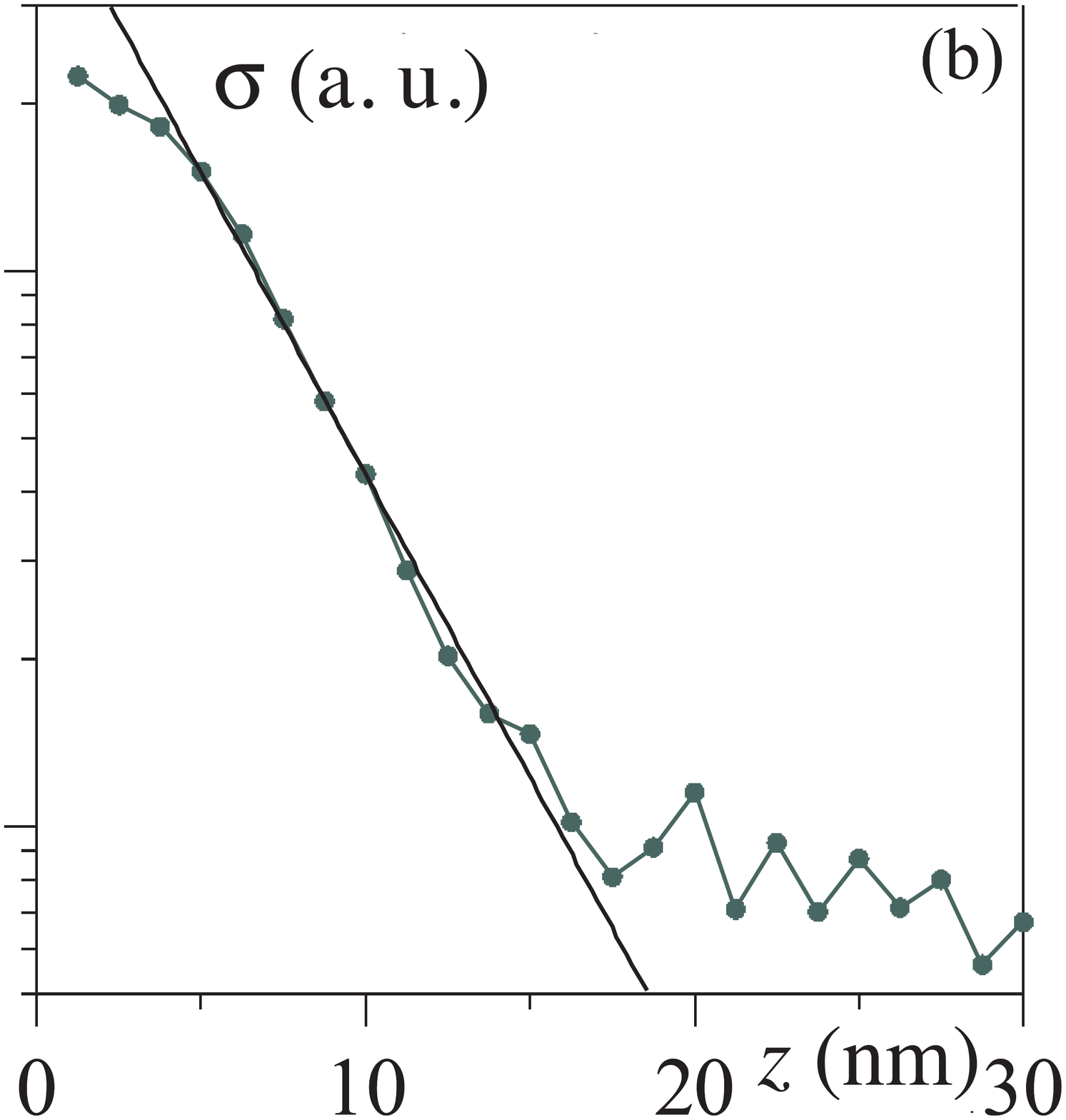}
  \includegraphics[width=0.315\linewidth]{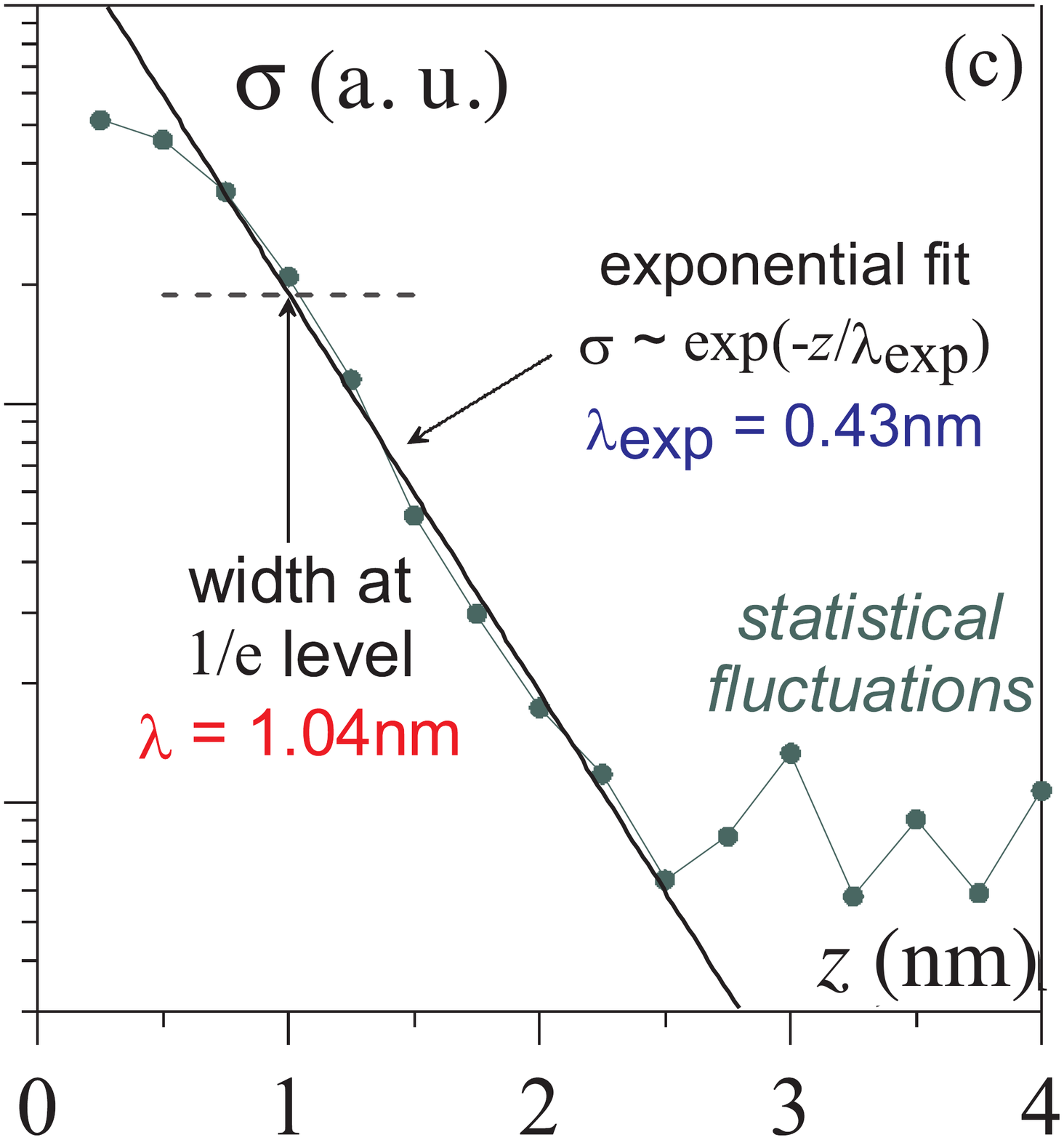}
  \vspace*{-14pt}
\end{center}
\caption{\label{charge-z}Distribution of the plasma charge (in arbitrary units)
over the longitudinal direction (log-linear plot).
The solid line represents the exponential fit, (a) $n_e = 10^{23}\mathrm{m}^{-3}$
(b) $n_e = 10^{25}\mathrm{m}^{-3}$, (c) $n_e = 10^{27}\mathrm{m}^{-3}$. In all cases $T = 1$eV.}
\end{figure}

\begin{figure}[h]  
\begin{center}
  \includegraphics[width=0.48\linewidth]{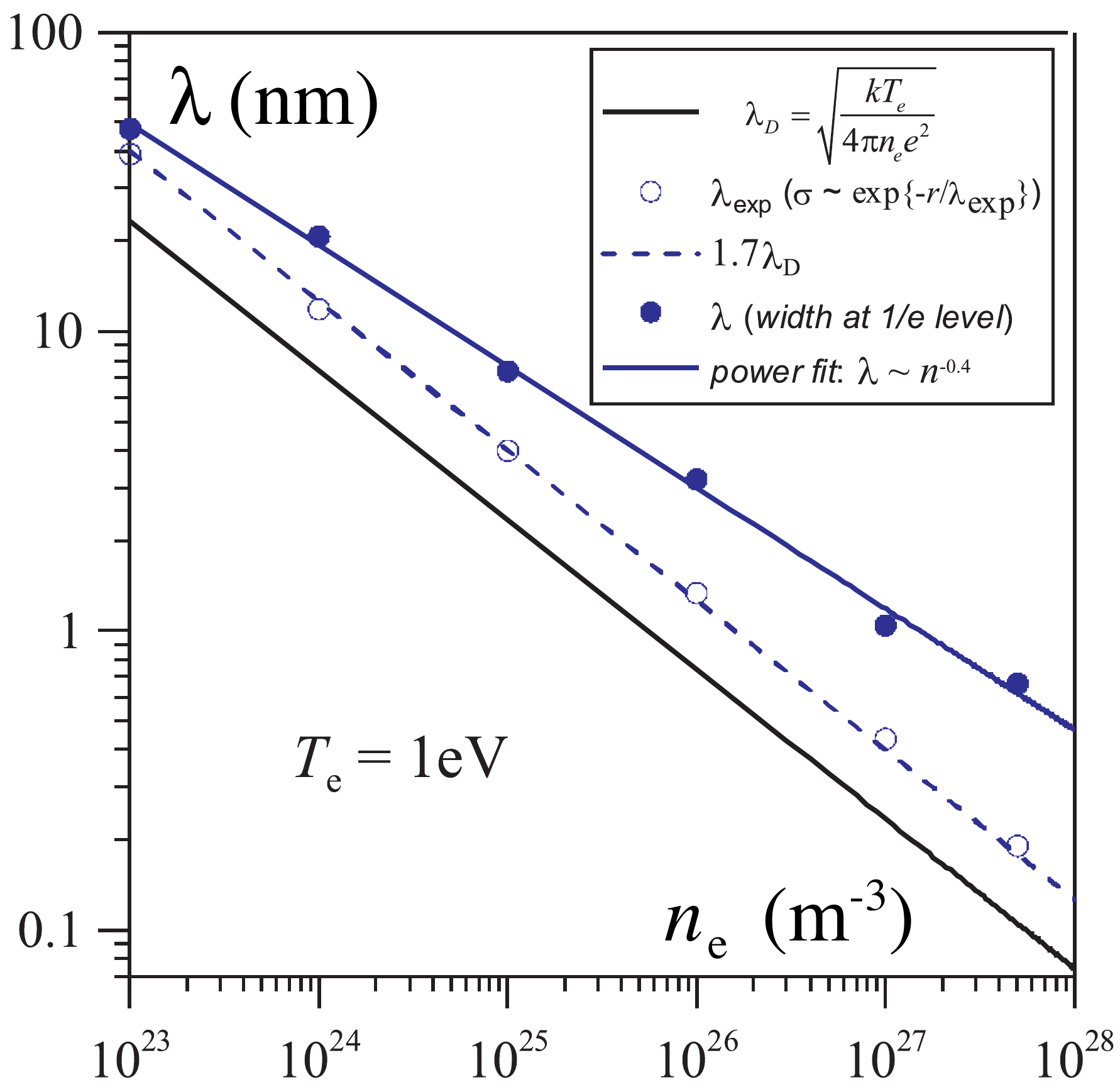}
  \includegraphics[width=0.48\linewidth]{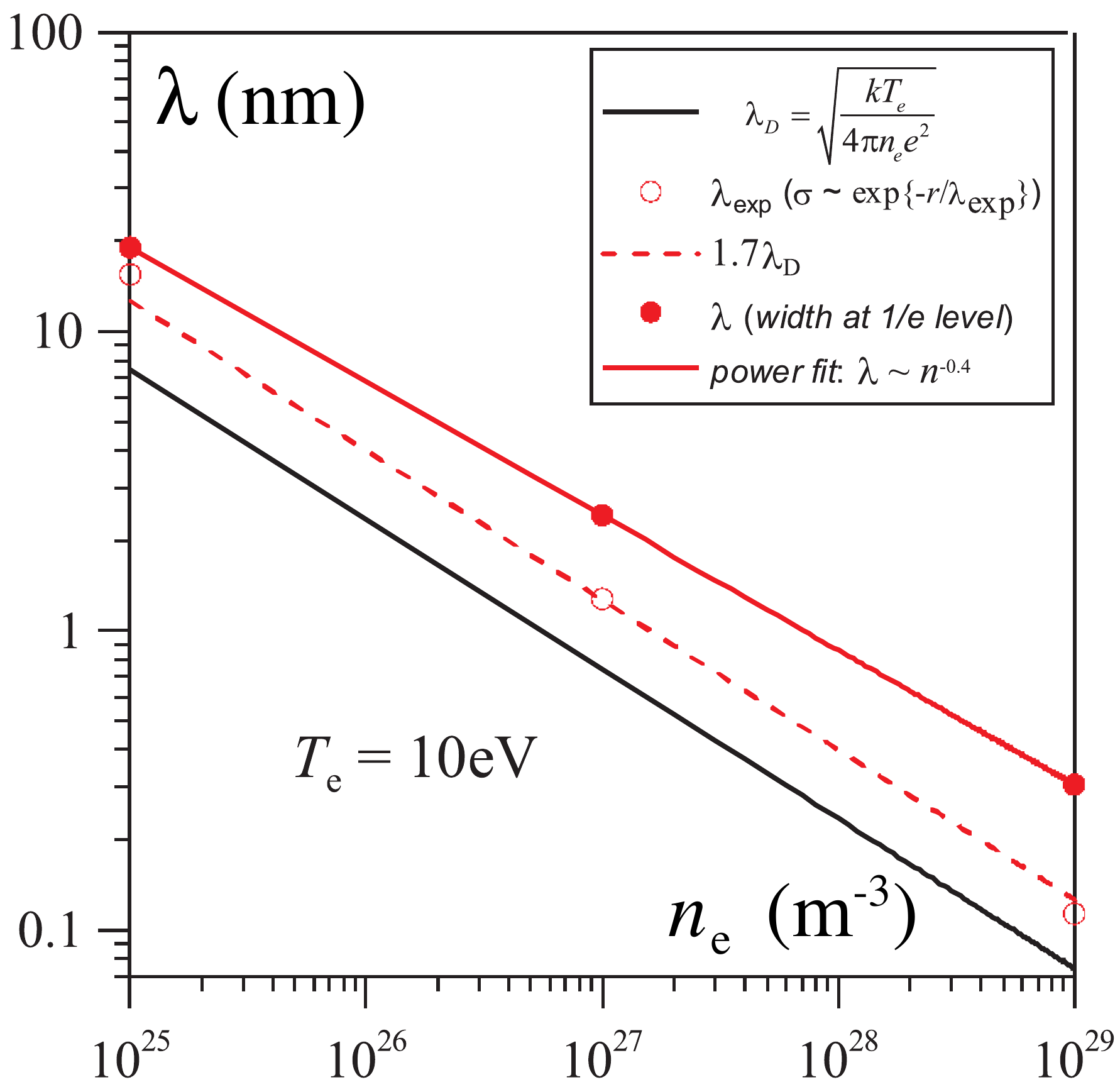}
  \vspace*{-14pt}
\end{center}
\caption{\label{lambda-n}The Debye length (lower solid line) and the sheath lengths
obtained from MD simulations depending on the electron number density.
Dashed line is related to the exponential fit
(see Fig.~\protect\ref{charge-z}), higher solid line represents the width $\lambda$ given
by the relation $\sigma(\lambda) = \sigma(0)/e$ where $\sigma(z)$ is the plasma charge
profile (Fig.~\protect\ref{charge-z}).
Left figure: $T = 1$eV, right figure: $T = 10$eV.}
\end{figure}

\begin{figure}[h]  
\begin{center}
  \includegraphics[width=0.50\linewidth]{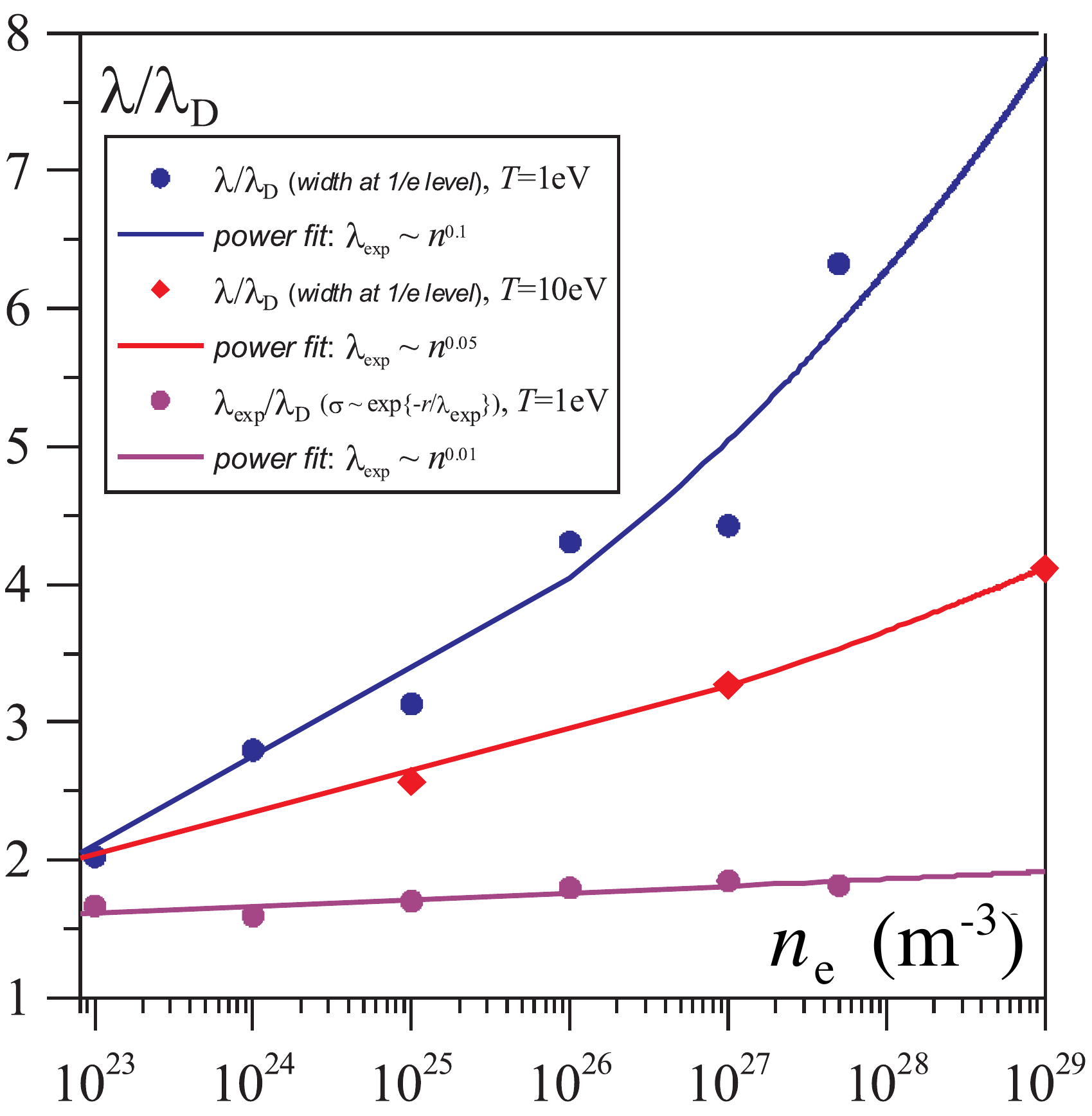}
  \includegraphics[width=0.48\linewidth]{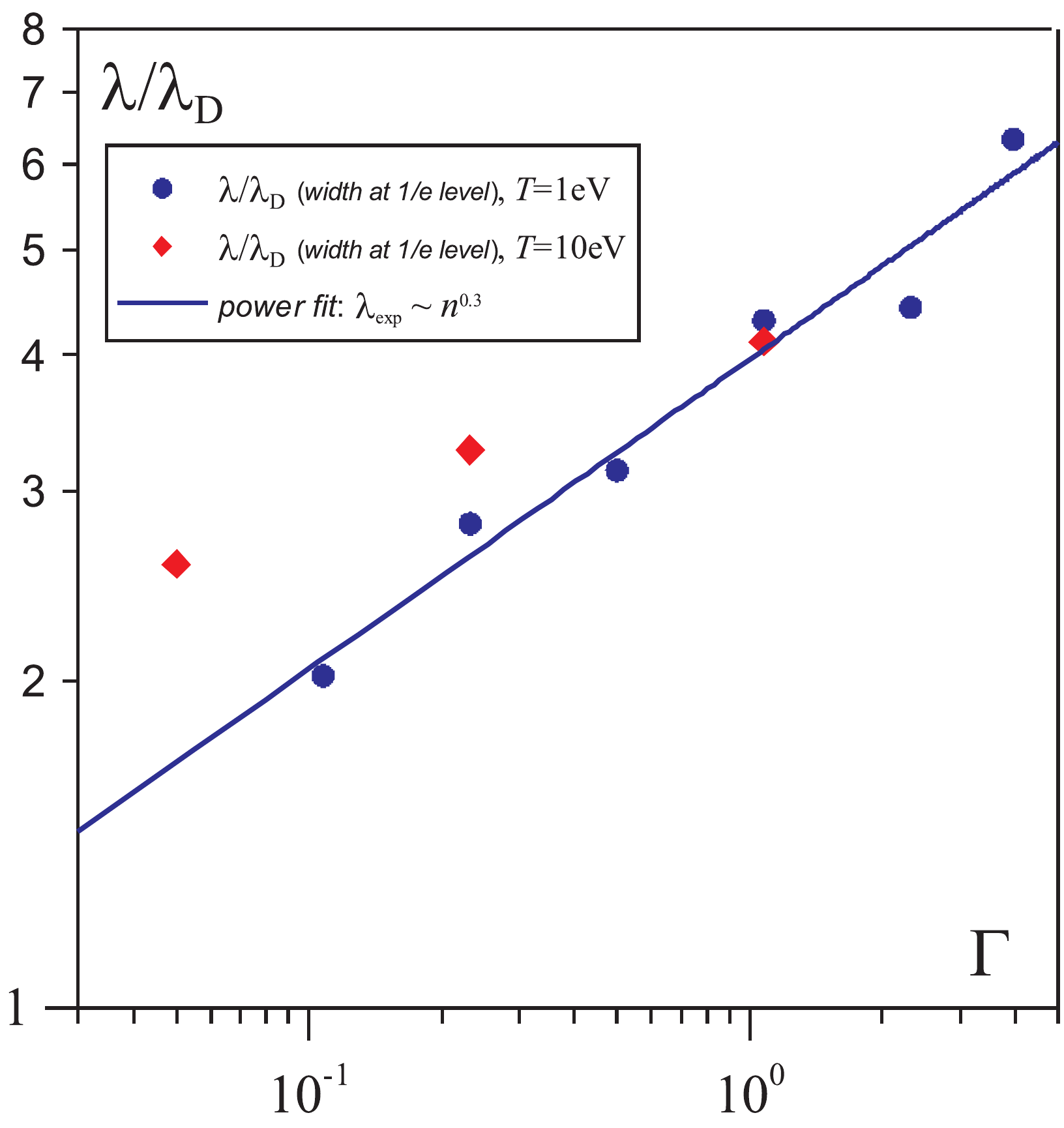}
  \vspace*{-14pt}
\end{center}
\caption{\label{lambdar-n}Dependence of the ratio between the sheath lengths obtained from MD
and the Debye length for different temperatures and plasma densities (see the legends).
Abscissa axis in plot (b) is the plasma nonideality parameter.}
\end{figure}

\begin{figure}[h]   
\begin{center}
  \includegraphics[width=0.49\linewidth]{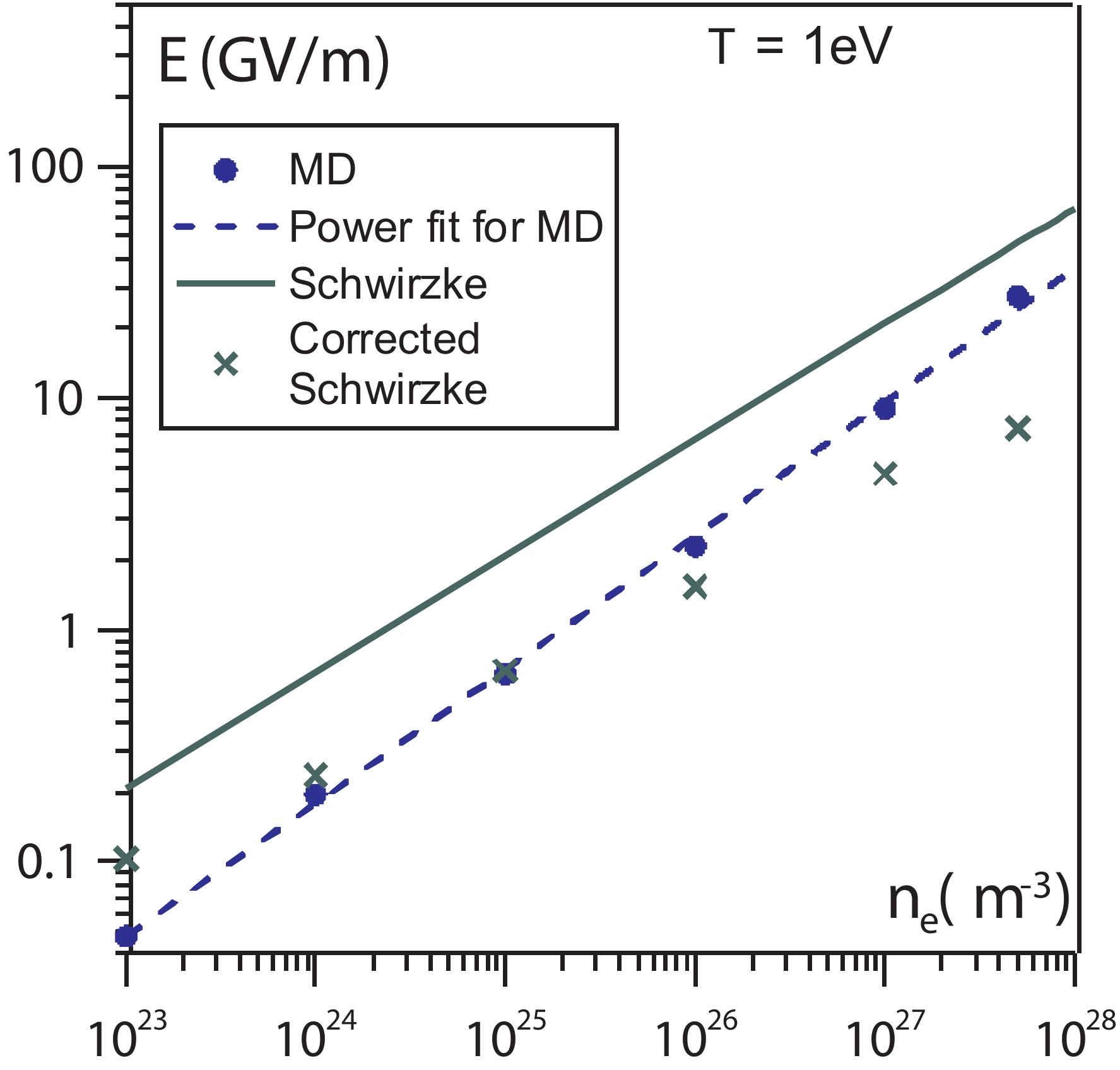}
  \includegraphics[width=0.49\linewidth]{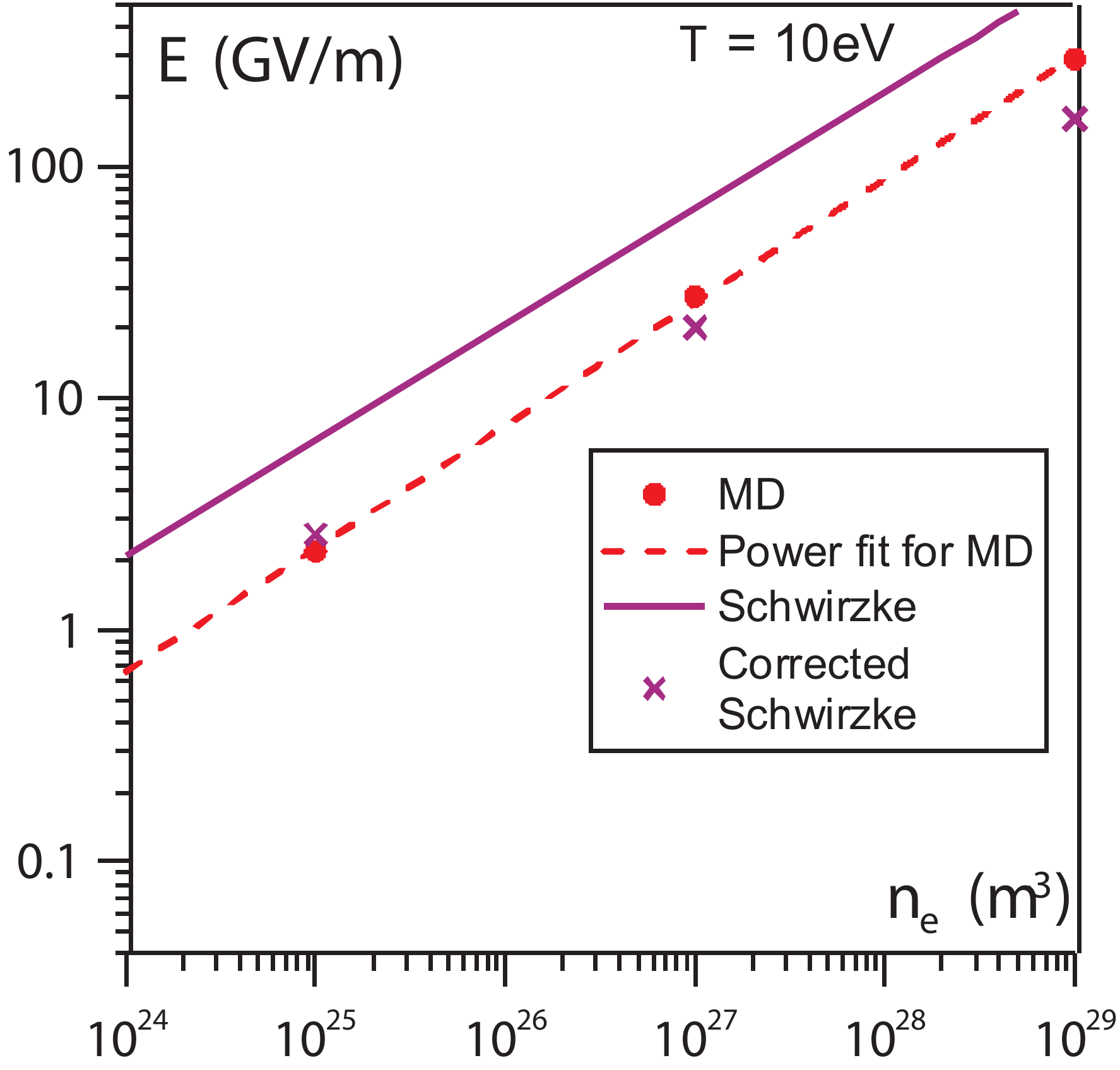}
  \vspace*{-14pt}
\end{center}
\caption{\label{surfiled-n}Dependence of the final surface electric field strength on the
electron number density for two values of temperature (shown on prot).
MD results are compared with the theoretical estimations from
\protect\cite{Schwirzke91}. Crosses correspond to the Eq.~(2) from~\protect\cite{Schwirzke91}
where the Debye length is replaced by the sheath lengths $\lambda$ obtained from MD
(Fig.~\protect\ref{lambda-n}).}
\end{figure}

\begin{figure}[h]  
\begin{center}
  \includegraphics[width=\myfigwidth]{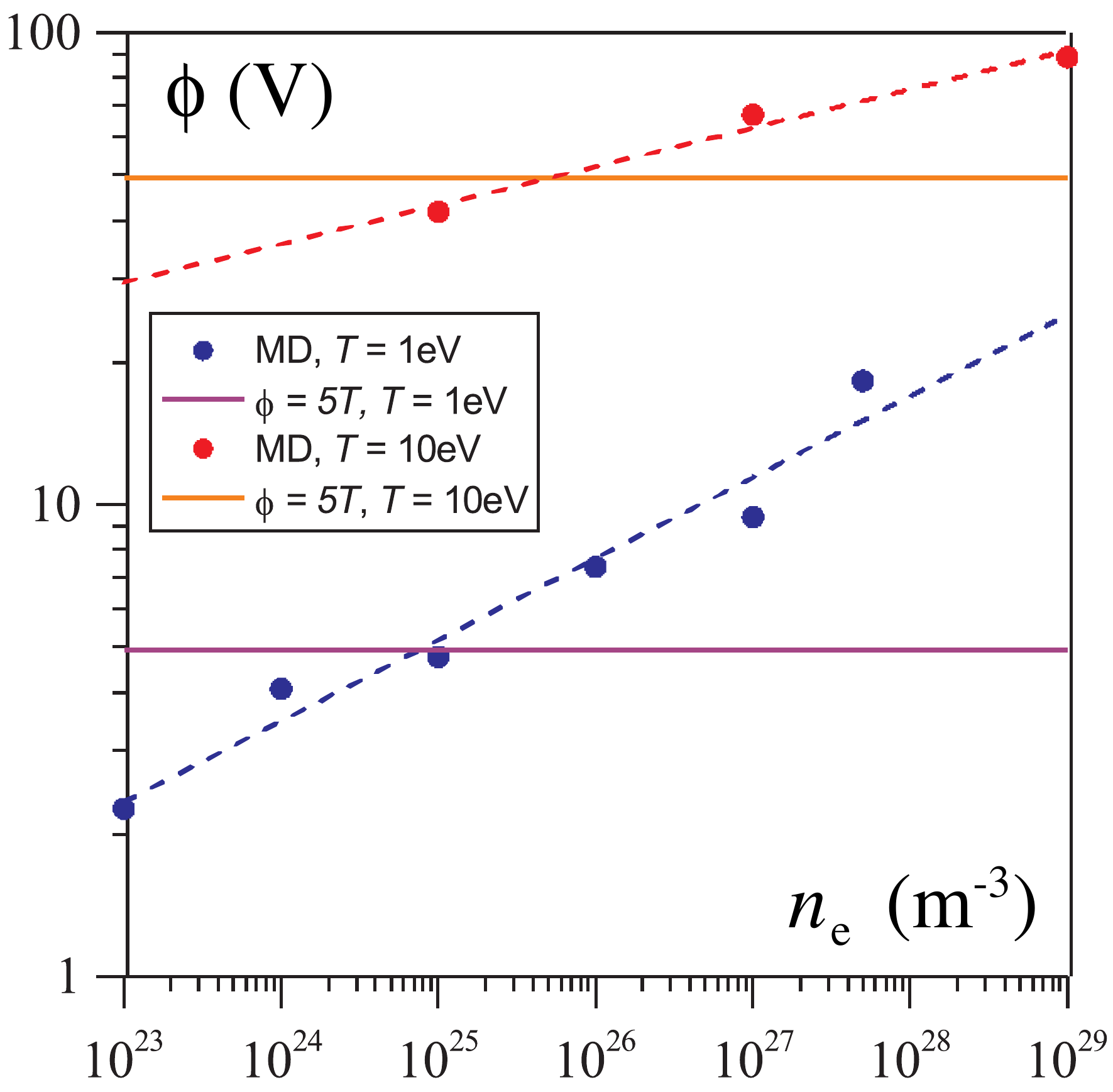}
  \vspace*{-14pt}
\end{center}
\caption{\label{plasmapot-n}
Plasma potential depending on the density for different temperatures. MD results
are compared with the theoretical estimations from \protect\cite{Schwirzke91}.}
\end{figure}

\begin{figure}[h]  
\begin{center}
  \includegraphics[width=\myfigwidth]{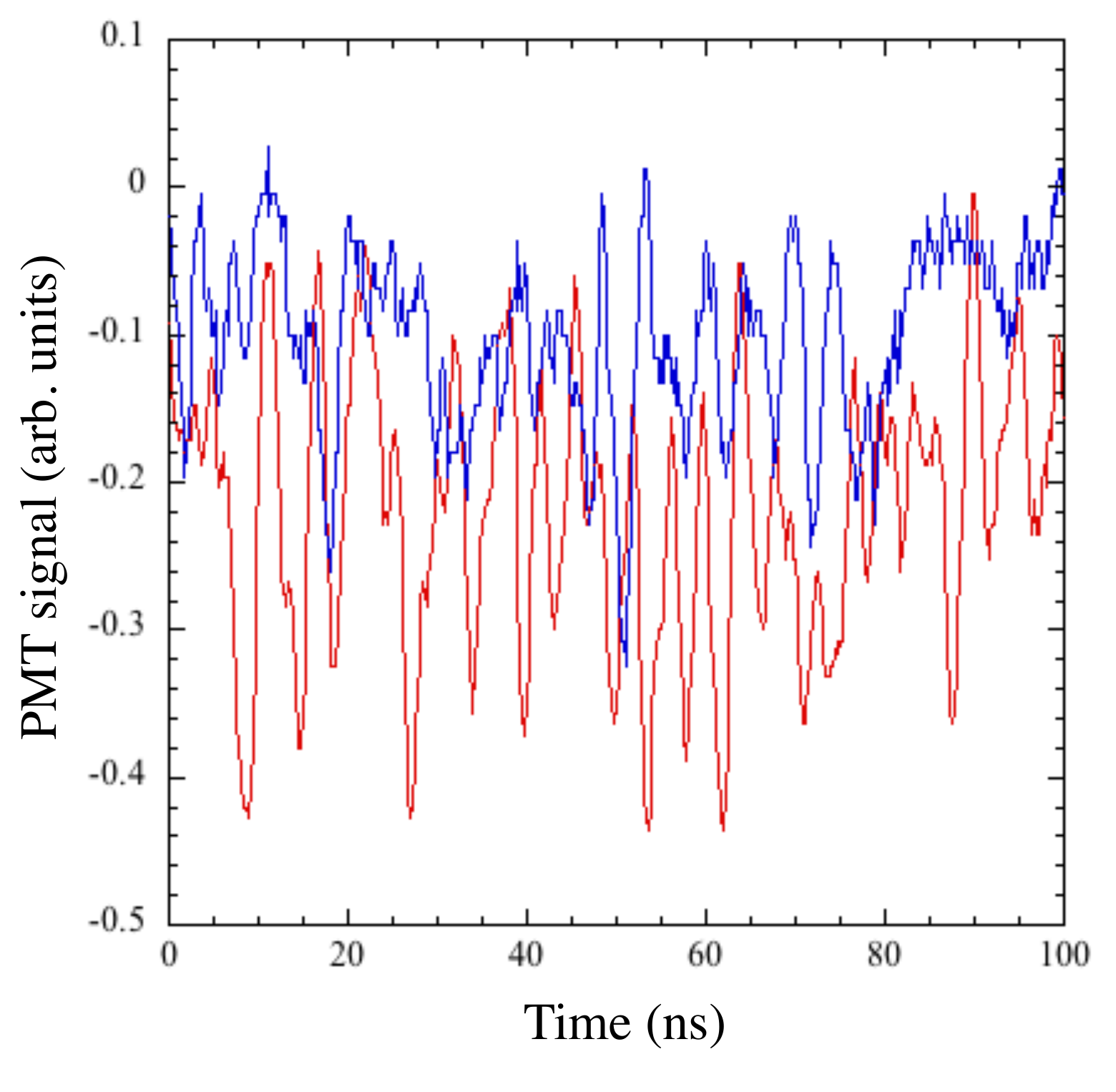}
  \vspace*{-14pt}
\end{center}
\caption{\label{PMT} The time evolution of visible light from an rf arc in an accelerator cavity during breakdown.  The red trace shows the conditions with $B=0$ and the blue trace shows the trace with a solenoidal field of $ B=3$ T.   Many experiments have seen similar behavior \cite{Anders}}.
\end{figure}

\begin{figure}[h]   
\begin{center}
  \includegraphics[width=\myfigwidth]{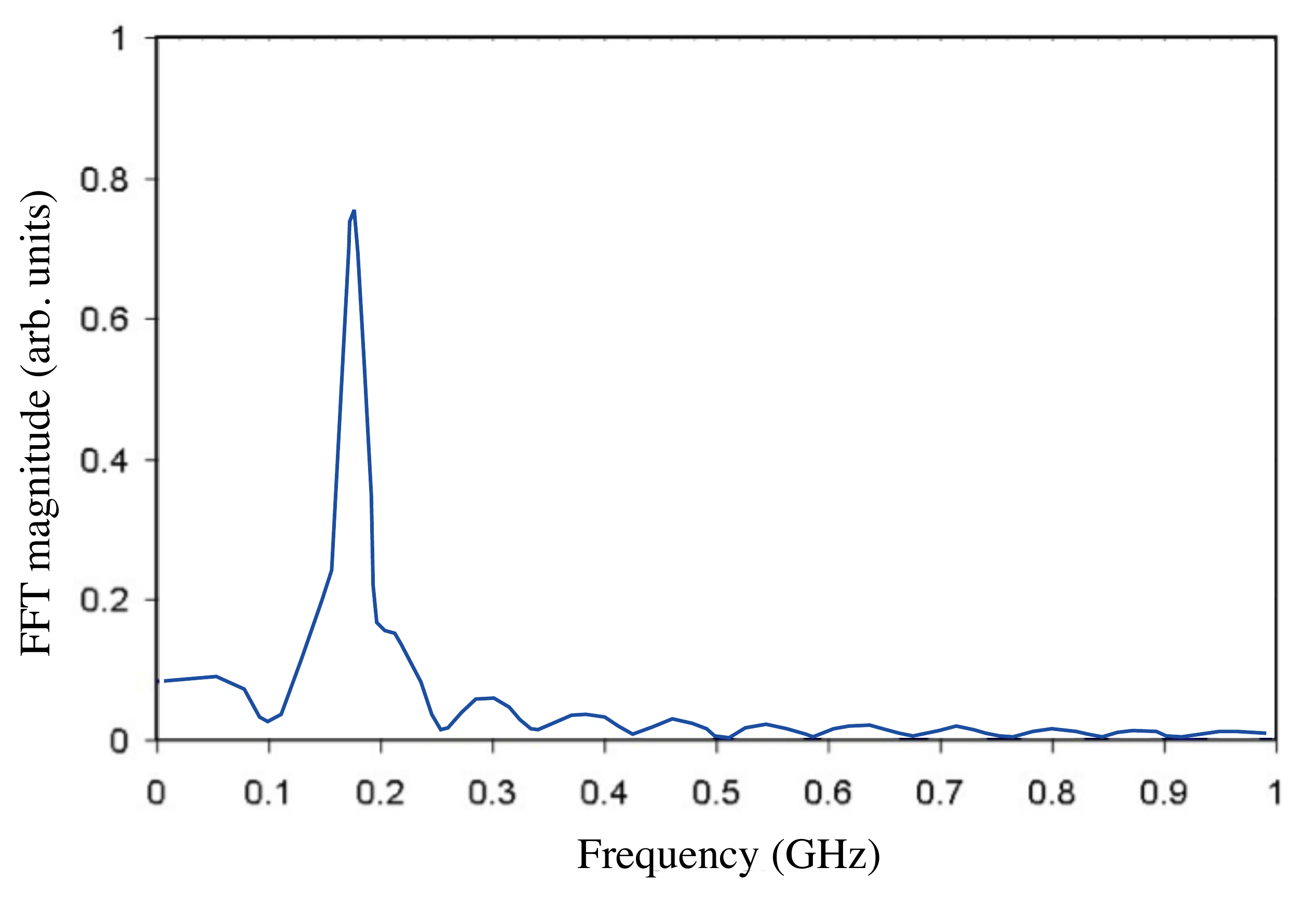}
  \vspace*{-14pt}
\end{center}
\caption{\label{FFT}
A Fast Fourier Transform of the B = 0 data in the previous Figure.  The peak around 200 MHz is similar to FFT plots of instabilities by Anders \cite{Anders}. }
\end{figure}

\begin{figure}[h]   
\begin{center}
  \includegraphics[width=\myfigwidth]{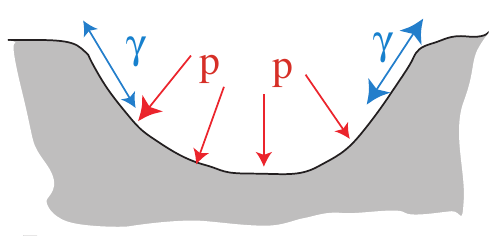}
  \vspace*{-14pt}
\end{center}
\caption{\label{p-gamma}
Plasma pressure, $p$ (red arrows), is opposed by the surface tension 2$\pi\gamma$r forces in the liquid metal.}
\end{figure}

\begin{figure}[h]    
\begin{center}
  \includegraphics[width=\myfigwidth]{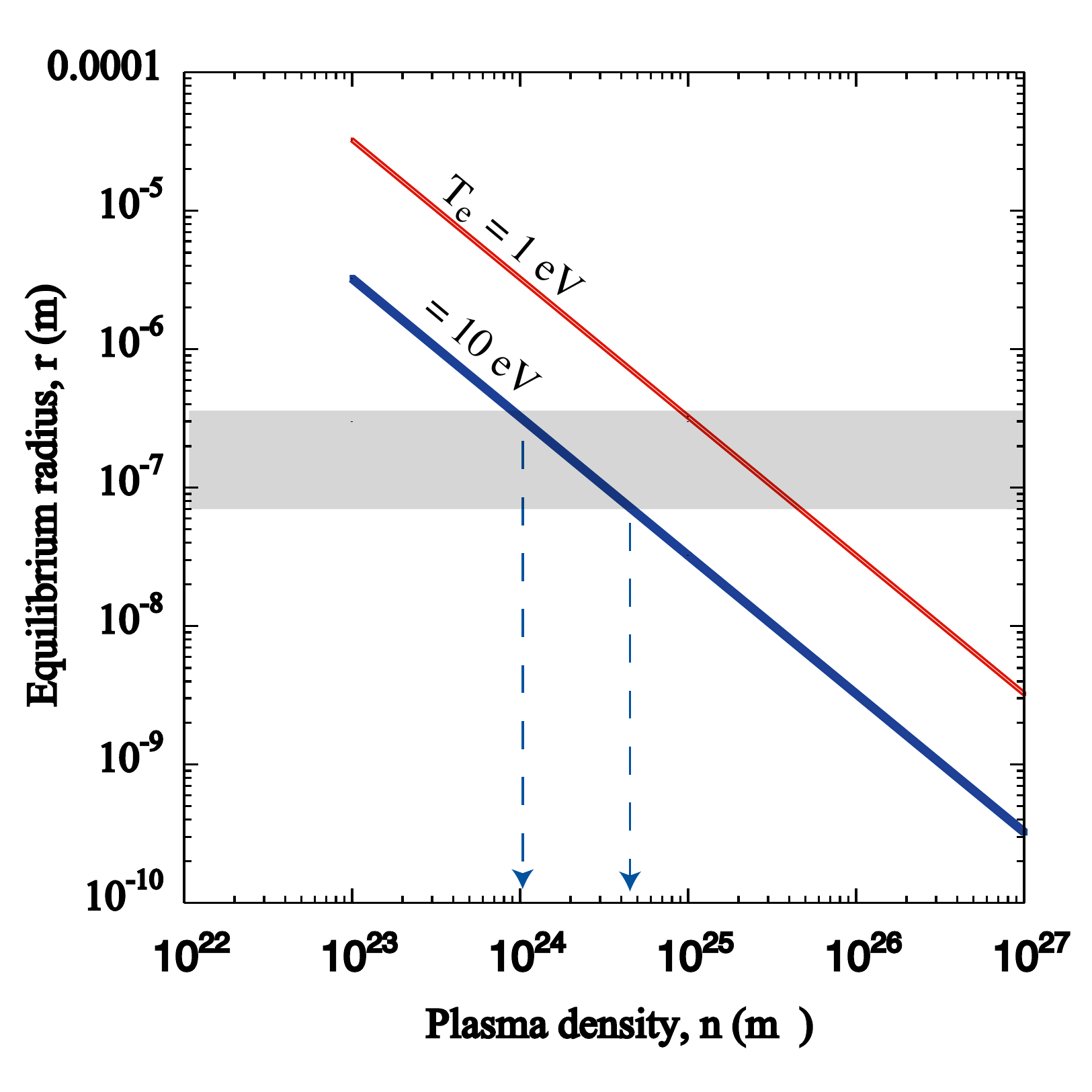}
  \vspace*{-14pt}
\end{center}
\caption{\label{r-n}
Equilibrium radius from surface tension and plasma pressure for two electron temperature plasmas compared with dimensions from ref \cite{Schwirzke91} and other measurements in the range ($\sim$80 - 350 $\mu$m), see Fig 14.}
\end{figure}

\begin{figure}[h]   
\begin{center}
  \includegraphics[width=\myfigwidth]{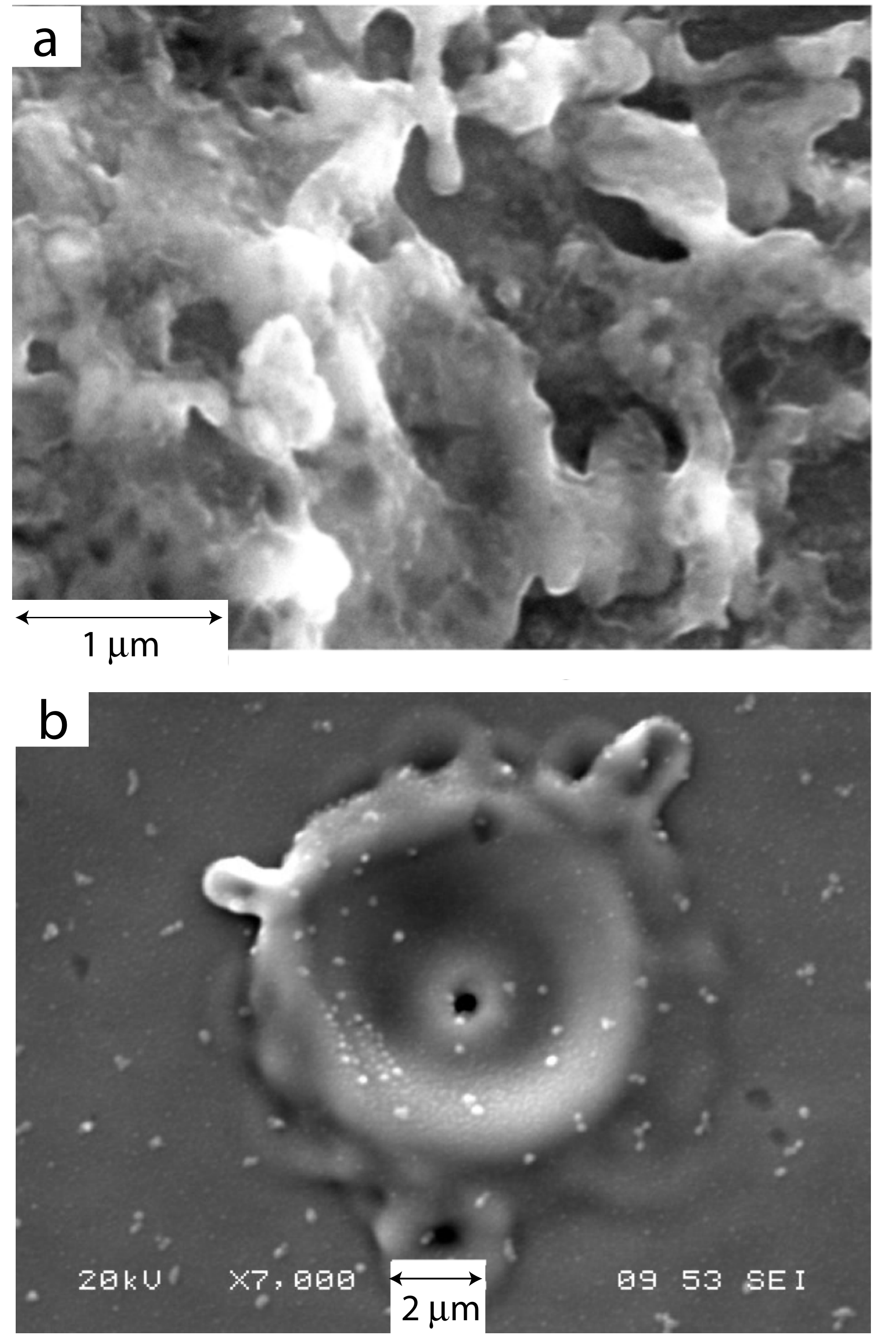}
  \vspace*{-14pt}
\end{center}
\caption{\label{SEM}
a) SEM image of unipolar arc tracks from a 201 MHz cavity coupler, showing considerable structure below 1 $\mu$m., b) Image of unipolar arc damage from
Castano \cite{castano}.}
\end{figure}


\begin{thebibliography}{99}

\bibitem{earhart} R. F. Earhart, Phil. Mag. {\bf 1,} 147 (1901).

\bibitem{Anders} A. Anders, {\it Cathodic Arcs, From Fractal Spots to Energetic Condensation}, Springer, New York (2008), Chapter 3.

\bibitem{Juttner} B. J\"{u}ttner, J. of Phys. D:  Appl. Phys., {\bf 34}, (2001) R1-3-R123.

\bibitem{mesyats} G. Mesyats, IEEE Trans. Plasma Sci., {\bf 23} (1995) 879.

\bibitem{kajita} S. Kajita, N. Ohno, S. Kakamura, J. Nuc. Mat. {\bf 415} (2011) 542.

\bibitem{CLIC} http://clic-study.web.cern.ch/clic-study/.

\bibitem{MAP} http://map.fnal.gov/

\bibitem{ITER} http://www.iter.org/mach

\bibitem{NoremPAC09}  Z. Insepov, J. Norem, D. Huang, S. Mahalingam, S. Veitzer, Proceedings of PAC09, Vancouver, B. C., Canada, May 4-8, (2009) 800.

\bibitem{NoremLINAC10}  Z. Insepov, J. Norem, T. Proslier, S. Mahalingam, S. Veitzer, Proceedings of LINAC10, Tsukuba Japan, Sept 12-17, (2010) 205.

\bibitem{Noremrf2011} Z. Insepov, J. Norem, S. Veitzer, S. Mahalingam, Proceedings of RF2011, June 1-3, Newport R. I. (to be published), arXiv:1108.0861.

\bibitem{InsepovNorem11} Z.~Insepov, J.~Norem, Th.~Proslier, A.~Moretti D.~Huang,
  S.~Mahalingam, S.~Veitzer; arXiv:1003.1736v3 (2010).

\bibitem{Schwirzke91} F.R.~Schwirzke;
  IEEE T.\ Plasma Sci.\ \textbf{19}, No. 5, 690 (1991).

\bibitem{Zwicknagel-JPA03} T.~Pschiwul, G.~Zwicknagel; J.\ Phys.\ A, \textbf{36},
  6251 (2003).

\bibitem{JETP05} I.V.~Morozov, G.E.~Norman; JETP, \textbf{100}, 370 (2005).

\bibitem{Murillo} L.X.~Benedict, J.N.~Glosli, D.F.~Richards, F.H.~Streitz,
  P.~Hau-Riege, R.A.~London, F.R.~Graziani, M.S.~Murillo, J.F.~Benage;
  Phys.\ Rev.\ Lett. \textbf{102}, 205004 (2009).

\bibitem{Donko} Z.~Donko; J.\ Phys.\ A, \textbf{42}, 214029 (2009).

\bibitem{Rostock-CPP09}
  T.~Raitza, H.~Reinholz, G.~R\"opke, I.~Morozov, E.~Suraud;
  Contrib.\ Plasma Phys.\ \textbf{49}, 498 (2009).

\bibitem{Suraud06}
  M. Belkacem, F. Megi, P.-G. Reinhard, E. Suraud, and G. Zwicknagel;
  Eur.\ Phys.\ J.\ D \textbf{40}, 247 (2006).

\bibitem{Zwicknagel-CPP03} G.~Zwicknagel, T.~Pschiwul;
  Contrib.\ Plasma Phys.\ \textbf{43}, 393 (2003).

\bibitem{CKelbg}
  A.V.~Filinov, M.~Bonitz, W.~Ebeling;
  J.\ Phys.~A \textbf{36}, 5957 (2003).

\bibitem{WPMD-JPA09} I.V.~Morozov, I.A.~Valuev;
  J.\ Phys.\ A, \textbf{42}, 214044 (2009).

\bibitem{velocityVerlet} M.P Allen, D.J. Tildesley, {\ it Computer simulation of liquids}, Oxford University Press Inc., NY, 1987.

\bibitem{MolSimul05} A.Y.~Kuksin, I.V.~Morozov, G.E.~Norman, V.V.~Stegailov,
  I.A.~Valuev; Mol.\ Simulat.\ \textbf{31}, 1005 (2005).

\bibitem{LangevinTherm} G.~S.~Grest and K.~Kremer,
Phys. Rev. A, {\bf 33}, (1986) 3628.

\bibitem{EKKR_book} W.~Ebeling, W.D.~Kraeft, D.~Kremp, G.~Ropke,
{\it Quantum Statistics of Charged Particle Systems}, Springer, New York (2005), Section~2.9.

\bibitem{Jackson_book} J.~D.~Jackson, \textit{Classical Electrodynamics}, Wiley, New York
(1999).

\bibitem{dolan} W. W. Dolan, Phys. Rev. , {\bf 91}, 510, (1953).

\bibitem{NIMBselfsput} Z. Insepov, J. Norem, S. Veitzer, Nucl. Instrum. and Meth in Phys. Res. B, {\bf 268}, (2010) 642.

\bibitem{NIMBselfsput1} Z. Insepov, J. Norem, A. Moretti, Proceedings of PAC 2011,  New York, NY, USA, p895.

\bibitem{He} J. He, N. M. Miskovsky, P. H. Cutler and M. Chung, J. Appl. Phys. {\bf 68(4)}, (1990) 1475.

\bibitem{Shortley} G. Shortley and D.Williams, {\it Elements of Physics: for Students of Science and Engineering}, Third Edition, Prentice-Hall, Inc., Englewood Cliffs, N.J. 1961 p 401.

\bibitem{copper}  T. Matsumoto, H. Fujii, T. Ueda, M. Kamai and K. Nogi, Meas. Sci. Technol. {\bf 16}, (2005) 432.

\bibitem{Tolman49} R. C. Tolman, J. Chem. Phys. {\bf 17}, 333 (1949).

\bibitem{castano} C. H. Castano G., presented at "Workshop on  Unipolar Arcs", Argonne, Jan. 29 (2010), https://twindico.hep.anl.gov/indico/conferenceDisplay.py?confId=69

\end{thebibliography}
\end{document}